\documentclass[preprint,showpacs,preprintnumbers,amsmath,amssymb]
{revtex4}

\usepackage{graphicx}
\usepackage{dcolumn}
\usepackage{bm}
\usepackage{lipsum}
\usepackage{color}
\usepackage{amsmath}
\usepackage{enumerate}
\usepackage{appendix} 

\usepackage{subfig}
\usepackage{amssymb}
\usepackage{txfonts}
\usepackage{wasysym}

\begin{document}

\title{Arrested dynamics of the dipolar hard-sphere model}

\author{Luis Fernando Elizondo-Aguilera$^1$, 
Ernesto Carlos Cort\'es-Morales$^2$, Pablo F. 
Zubieta Rico$^2$, Magdaleno Medina-Noyola$^2$,
Ram\'on Casta\~neda-Priego$^3$, Thomas Voigtmann$^{1,4}$ and 
Gabriel P\'erez-\'Angel$^5$}
\affiliation{$^1$ Institut f\"ur Materialphysik im Weltraum, Deutsches
Zentrum f\"ur Luft-und Raumfahrt (DLR), 51170 K\"oln, Germany,}
\affiliation{$^2$ Instituto de F\'isica, Universidad Aut\'onoma de San
Luis Potos\'i, Av.\ Manuel Nava 6, Zona Universitaria, 78290 San Luis
Potos\'i,
San Luis Potos\'i, M\'exico,}
\affiliation{$^3$ Departamento de Ingenier\'ia F\'isica, Divisi\'on
de Ciencias e Ingenier\'ias, Universidad de Guanajuato, Loma del Bosque
103, 37150 Le\'on, M\'exico.}
\affiliation{$^4$ Department of Physics, Heinrich-Heine-Universit\"at
D\"usseldorf, Universit\"atsstra\ss{}e 1, 40225 D\"usseldorf, Germany,}
\affiliation{$^5$ Departamento de F\'isica Aplicada, 
CINVESTAV del IPN, A.\ P.\ 73 ``Cordemex'', 97310 M\'erida, 
Yucat\'an, M\'exico,}

\date{\today}

\begin{abstract}
We report the combined results of molecular dynamics simulations and 
theoretical calculations concerning various dynamical arrest transitions 
in a model system representing a dipolar fluid, namely, $N$ (soft core) 
rigid spheres interacting through a truncated dipole-dipole potential. 
By exploring different regimes of concentration and temperature, we 
find three distinct scenarios for the slowing down of the dynamics of 
the translational and orientational degrees of freedom: 
At low ($\eta=0.2$) and intermediate ($\eta=0.4$) volume fractions, 
both dynamics are strongly coupled and become simultaneously arrested 
upon cooling. At high concentrations ($\eta\geq 0.6$), the translational 
dynamics shows the features of an ordinary glass transition, 
either by compressing or cooling down the system, but with the orientations 
remaining ergodic, thus indicating the existence of partially arrested 
states. In this density regime, but at lower temperatures, the relaxation 
of the orientational dynamics also freezes.  
The physical scenario provided by the simulations is discussed and 
compared against results obtained with the self-consistent generalized 
Langevin equation theory, and both provide a consistent description of 
the dynamical arrest transitions in the system. 
Our results are summarized in an arrested states diagram 
which qualitatively organizes the simulation data and provides a 
generic picture of the glass transitions of a dipolar fluid.
\end{abstract}

\pacs{xxxxxxxx yyyyyyyy}

\maketitle

\section{Introduction}

The substantial progress in the synthesis and fabrication of anisotropic
colloids \cite{glotzer,walther,zhangt,cayre}, along with their unique
ability for self-assembly and structural reconfiguration
\cite{walther,zhangt,safran,nych,butter}, provides a route for the design
of new specialized materials with novel physical properties and with specific
functionalities of high technological interest.
Colloids of non-spherical shapes have long been known, but recognition
that anisotropy in interactions constitutes another potential tool for
engineering particular targeted structures has brought a widespread
research enthusiasm in dipolar systems
\cite{safran,nych,butter,tlusty,klokkenburg,rovigatti,cattes,sindt,
koperwas,weis,belloni,yethiraj}.
Even at thermodynamic \emph{equilibrium} conditions and in the absence
of external fields, dipolar suspensions tend to assemble into 
energetically favorable \emph{head-to-tail} configurations and typically 
display a rich structural, dynamical and phase behavior arising from  
the complex and highly anisotropic nature of the dipolar interaction 
\cite{belloni,cattes,weis,pincus,blaak1,blaak2,goyal1,goyal2,goyal3}.

The current interest in structures of increasing complexity, however, 
has also triggered investigations of self-assembly under
\emph{non-equilibrium} conditions  \cite{testard,varrato,kim,stratford}.
Colloidal suspensions, in particular, have been observed to undergo 
distinct transitions to non-crystalline amorphous states. At high 
densities, for instance, they can form glassy states, where the main 
underlying physical mechanism for dynamical arrest is nearest-neighbor 
\emph{caging} inhibiting individual motion \cite{pusey,vanmegen}.
At low densities, they may also undergo gelation by increasing the mutual
interactions among particles, prompting long-lived physical (reversible)
bonding between particles, which thus facilitates the formation of
percolated networks capable of sustaining weak mechanical stresses
\cite{foffi,zacarelli,pastore,sciozacca}. Striking similarities, but also
fundamental differences, have been highlighted between the microscopic
dynamics and the mechanical response of both gels and glasses
\cite{pastore,zacarelli}.

Dipolar colloids have been the subject of numerous investigations
based on computer simulations \cite{rovigatti,dijkstra,rovigatti-PRL-2011,goyal1,goyal2,goyal3,blaak1,blaak2}. 
Goyal \emph{et al.}, for example,  carried out molecular dynamics 
(MD) studies aimed at outlining the \emph{equilibrium} phase diagram 
in the packing fraction ($\eta$) \emph{vs} temperature ($T$) plane of 
monocomponent fluids of hard spheres (HS) with embedded dipoles 
\cite{goyal1} and of binary mixtures with difference in size and dipolar 
moment \cite{goyal2,goyal3}. One of the most interesting aspects of such 
diagrams is the observation that, besides different crystalline  
(\emph{e.g.,} fcc, hcp, bct), fluid, and string-fluid \emph{equilibrium} 
phases, a region of percolated networks of crossed-links chains exists, 
occurring at intermediate temperatures and at nearly all the volume 
fractions considered, thus describing a region of gel-like states.  
In addition, Blaak \emph{et.al.,} \cite{blaak1,blaak2} demonstrated 
that, at low densities ($\eta<0.1$), a slight elongation of the 
particles into dumbbells favors branching of dipolar chains, and hence, 
space-spanning networking towards the gel state.
It was also highlighted that this behavior is accompanied by a
noteworthy slowing down of the translational dynamics, closely resembling
well known features of physical (reversible) gelation in systems with
short-ranged but spherically-symmetric depletion attractions 
\cite{sciozacca}, just as it occurs in the case of 
colloid-polymer mixtures \cite{pham,bergenholtz} and systems with 
adhesive (sticky) interactions \cite{ramon1,ramon2}.

The theoretical modeling of the dynamical arrest of dipolar fluids is not 
abundant. Schilling and Scheidsteger \cite{schilling-PRE-1997} applied 
the mode coupling theory (MCT) of the glass transition (GT) to predict 
the existence of several dynamical arrest transitions in a dipolar 
HS fluid and outlined the corresponding \emph{non-equilibrium} states 
diagram. More recently, the same essential features were also observed within 
the self-consistent generalized Langevin equation (SCGLE) theory 
\cite{Elizondo-PRE-2014}. To date, however, characterizations of the 
glassy and gel behavior of dipolar liquids, based on the simultaneous use 
of simulations and theoretical approaches, are rather scarce. Thus, it 
remains to be understood whether the competition between caging and
bonding, along with the highly anisotropic character of the interactions
(which couples translational and orientational motion)
mediates the various dynamical arrest transitions. In particular, the 
role of the orientational dynamics in both the glass and gel transitions
has not been fully determined.

A collection of $N$ rigid spheres of diameter $\sigma$ bearing a
dipolar moment $\bm{\mu}$ is clearly one of the simplest representations
of a dipolar fluid, a model that possesses a remarkable theoretical 
significance \cite{wertheim,gray-molec-fluid-1984}, since it combines both 
spherical entropic and anisotropic energetic interactions. 
Notwithstanding its apparent conceptual simplicity, the distinct glassy 
behavior of such system is far from being completely understood. Hence, 
a main aim of this work is to report the results of extensive MD 
simulations carried out on a slightly simplified version of 
this model,  introduced for practical convenience. 
We refer to a system of spherical particles whose repulsive-core 
interaction is given by the so-called Weeks-Chandler-Andersen (WCA) 
potential \cite{weeks-JCP-1971}.
For simplicity, and in order to focus on the effect of anisotropy in 
the interactions, rather than on their long-range nature, the particles 
further interact by a truncated dipole-dipole potential. 
For this model, we have investigated different regimes of concentration
and temperature where, on the basis of previous theoretical
\cite{schilling-PRE-1997,Elizondo-PRE-2014} and simulation
\cite{goyal1,goyal2,goyal3,blaak1,blaak2} results, various dynamical
 arrest transitions, including gels and glasses, are expected to occur.

Therefore, instead of considering \emph{equilibrium} phases, in this 
contribution we are specifically interested in investigating those
morphological transformations driven by conditions of dynamical
arrest. In particular, we find three types of transitions:
\emph{(i)} the simultaneous arrest of the dynamics of both the translational
(TDF) and orientational degrees of freedom (ODF), occurring at low and
intermediate concentrations upon lowering the temperature; 
\emph{(ii)} an ordinary GT for the translational dynamics, occurring at 
high densities and high-to-intermediate temperatures, but where the 
relaxation of the orientational dynamics remains ergodic; and \emph{(iii)} 
the subsequent arrest of the ODF in a positionally frozen structure of 
particles, occurring also at high densities but lower temperatures.
Our findings thus suggest the existence of at least three different GTs 
in the parameters space $(\eta,T^*)$, which describe distinct microscopical 
underlying mechanisms for arrest.

Besides the analysis of the dynamics of the TDF, given in terms 
of observables such as the self-part of the Fourier transform of the 
van Hove correlation function and the mean square displacement, we 
also consider explicitly the dynamics of the ODF of the model towards the 
GT. This is described in terms of a proper set of orientationally sensitive 
correlation functions 
\cite{wertheim,gray-molec-fluid-1984,schilling-PRE-1997,Elizondo-PRE-2014} 
and the corresponding angular mean square deviations. 
As we show below, this allows us to unravel important features of the 
dynamical arrest transitions of a dipolar fluid not fully described in 
previous investigations.

Finally, to provide a stronger physical insight, we complement the 
MD simulation results with a theoretical analysis based on the 
self-consistent generalized Langevin equation (SCGLE) theory 
\cite{Elizondo-PRE-2014} of dynamical arrest.
We particularly focus on the study of the so-called non ergodicity
parameters, describing the long-time dynamics of the fluid in the 
vicinity of each glass transition, and summarized in an arrested state 
diagram, which organizes qualitatively our results from simulations. 
The physical scenario emerging from the theory is in remarkable 
agreement with the features observed in the simulations, thus providing 
a consistent physical description of dynamical arrest.
 
This paper is organized as follows:
in section \ref{simuldetails} we define the model for the MD simulations
and discuss the physical observables of interest along with the 
simulation details; in section \ref{simresults} we present and discuss 
the MD simulation results; in section \ref{Theoryanddiagrams} we 
complement these results with a theoretical discussion and show that 
they are consistent, which provides a comprehensive picture of the 
glassy behavior of the dipolar fluid. 
Finally, in section \ref{conclusions} we formulate our conclusions.

\section{Simulations: methodological aspects}\label{simuldetails}

The standard tools of MD cannot be directly applied to simulate a dipolar
hard-sphere (DHS) fluid due to the mismatch between the discontinuous HS
potential and the \emph{soft} dipolar contribution. The former can be
properly approached only via an \emph{event-driven} algorithm
\cite{alder-JCP-1959,lubachevsky-JCP-1991}, whereas the latter
requires a \emph{continuous-time} approach. One may use
a Monte Carlo (MC) sampling considering both \emph{hard} and
\emph{soft} potentials, but this only would be useful if our main
purpose were the determination of the static properties.
The absence of a well defined time variable in MC simulations, however, 
renders the evaluation of dynamical quantities complicated (MC time is
nevertheless used some times, see, for instance, Ref.
\cite{berthier-JPCM-2007}). To circumvent these limitations, we have 
introduced a set of simplifying modifications to the pair interactions 
between dipolar spheres. Similarly, rather than simulating an exhaustive 
catalog of physical properties, here we shall focus on the dynamical 
properties associated with the (translational and rotational) Brownian 
motion and tracer diffusion properties, as explained in what follows.

\subsection{Model system.} \label{model}

We have simulated a system of $N$ spherical particles in a volume $V$, 
with a pairwise potential of interaction between particles (labeled as 
$a$ and $b$) given by

\begin{equation}
    U^{ab}({\bf r}_a, \bm{\mu}_a; {\bf r}_b,  \bm{\mu}_b) =
    U_\text{WCA}^{ab} (|{\bf r}_b - {\bf r}_a|) +
    U_\text{TD}^{ab}({\bf r}_b - {\bf r}_a, \bm{\mu}_a,
    \bm{\mu}_b).\label{potential}
    \end{equation}

\noindent where ${\bf r}_a$ and ${\bf r}_b$ are the vectors describing
the position of the center of mass of particles $a$ and $b$, respectively, 
and with $\bm{\mu}_a$ and $\bm{\mu}_b$ being their corresponding 
dipolar moments. 
The first term on the right side of Eq. \eqref{potential} represents the 
hard-core interaction, and is given by the so-called 
Weeks-Chandler-Andersen (WCA) potential \cite{weeks-JCP-1971}

    \begin{eqnarray}
    U_\text{WCA}^{ab} (r_{ab}) &=& 4 \epsilon_{\text{\tiny{WCA}}}^{ab} \left[
    \left(\frac{\sigma_{ab}}{r_{ab}}\right)^{12} -
    \left(\frac{\sigma_{ab}}{r_{ab}}\right)^6
    + \frac{1}{4}
    \right] \qquad \text{for}\; r_{ab} < 2^{1/6} \sigma_{ab}, \nonumber\\
    &=& 0 \qquad \text{otherwise,}\label{wcapotential}
    \end{eqnarray}

\noindent where $r_{ab}\equiv|{\bf r}_{ab}|=|{\bf r}_b - {\bf r}_a|$ and
the particles have sizes $\sigma_a$ and $\sigma_b$, taken from a
distribution with an average $\sigma_{ave} = 1$ (this last equality sets
the length  scale of the model) and we define $\sigma_{ab} = (\sigma_a +
\sigma_b)/2$.
Since we want to study glassy behavior at different densities, we have 
considered an equimolar binary mixture using two sizes 
$\sigma_1=\sigma_{ave}+\Delta\sigma$ and $\sigma_2=\sigma_{ave}
-\Delta \sigma$
The deviation in sizes has been set
to $\Delta\sigma = 1/6$, which thus gives a ratio $\sigma_1/\sigma_2 =
1.4$, a common size ratio used to avoid crystallization in dense HS
fluids \cite{berthier-PRE-2009}.

The second term on the right side of Eq. \eqref{potential} represents
the anistropic dipolar contribution to the interaction. In order to avoid
all the complexities inherent to the specialized treatment of long-ranged 
($\sim1/r^3$) interactions \cite{allen-tildesley}, we have considered a 
truncated dipolar potential which may be written as

    \begin{eqnarray}
    U_\text{TD}^{ab}({\bf r}_{ab}, \hat{\bm{\mu}}_a, \hat{\bm{\mu}}_b)
    &=&
	\epsilon_{\text{\tiny{DIP}}}^{ab}\left[ 
    \frac{1}{r_{ab}^3} -
    \frac{1}{R^3} +
    3 \frac{r_{ab} - R}{R^4}
    \right] \times \nonumber \\
    && \left[
    \hat{\bm{\mu}}_a \cdot \hat{\bm{\mu}}_b - 3 (\hat{\bm{\mu}}_a \cdot
    \hat{\bf r}_{ab} )
    (\hat{\bm{\mu}}_b \cdot \hat {\bf r}_{ab}) \right]
    \qquad \text{for}\; r_{ab} < R, \nonumber \\
    &=& 0
    \qquad \text{otherwise}.\label{dipdip}
    \end{eqnarray}
    
Here ${\bf \hat r}_{ab} = {\bf r}_{ab} / r_{ab}$,
$\hat{\bm{\mu}}_a\equiv\bm{\mu}_a/\mu_a$, $R$ is a
\emph{cut-off} distance for the dipolar interaction and
$\epsilon_{DIP}^{ab}=\mu_a\mu_b$. 
In this work we are setting $R = 3 \sigma_{ave}$.
The magnitude of a given dipole moment is proportional to its diameter 
to the power $3/2$, that is, 
$\mu_i = \mu_0(\sigma_i/\sigma_{ave}) ^{3/2}$ ($i=1,2$), 
where $\mu_0$ is a common dipole scale taken here as $\mu_0 = 1$.
The reason to introduce this dependence is to keep the interaction
at contact (\emph{i.e.,} for $r_{ab} \approx \sigma_{ab}$) approximately
the same for all contacts, be they  between large dipoles, small
dipoles, or a small dipole with a large one; otherwise the attraction
at contact between small dipoles is substantially larger than the one
between large dipoles, and segregation phenomena may appear.
For simplicity all dipoles are assumed to have mass $m = 1$ (this
sets the mass scale) and moment of inertia $I = 1/10$, and the scale of 
the WCA potential is set by $\epsilon_{\text{\tiny{WCA}}}^{ab}=1$
for all $a,b=1,2$, fixing in this way both energy and time scales.

\subsection{Physical observables.} \label{observables}

In this contribution we focus on the single particle dynamics, which is 
sampled with better statistics than collective dynamics. Specifically, 
we consider the \emph{self}-part of the generalized intermediate scattering 
functions (ISFs) and  mean square deviations defined below as a measure of
translational and rotational diffusion. To monitor the stability of the 
simulations, we have also checked  translational and rotational kinetic 
energies, potential energies, and pressure.

Also reviewed, and of special interest, are those quantities that may
signal a breakdown of the homogeneity or the isotropy of the fluid.
To test for possible crystallization, we have used the
$q_4, q_6, Q_4$ and $Q_6$ orientational order parameters defined by
Steinhardt and Nelson \cite{steinhardt-PRB-1983,tenvolde-JCP-1996}.
To look for possible segregation, instead, we introduce a simple
"\emph{sameness}" parameter, which measures how many dipoles of the same
size are within a cut-off distance of any given dipole, compared to how
many dipoles of the other size can be found.
More concretely, this is defined for the $a$-th dipole as
$S_a = \sum_{\langle nn\;a \rangle} z_{ab}/N_{\langle nn\;a\rangle}$, 
where we take $z_{ab} = 1$ if dipoles $a$ and $b$ are of the same size, and
$z_{ab} = -1$ otherwise. Here the sum is taken over all nearest neighbors
(\emph{nn}) to $a$, as defined by the cut-off distance, and 
$N_{\langle nn\;a\rangle}$ is the number of those nearest neighbors. It is 
also quite important to check the variance in $S$, since a small value for 
this variance indicates some degree of ordering. The cutoff distance used 
here is the same one used for the aforementioned Steinhard-Nelson order 
parameters, and has been taken for our purposes as the distance for the 
first minimum in the pair distribution function, $g(r)$, which is larger 
than $\sigma_1$. Finally, we also monitor the total magnetization, 
${\bf M} = \sum_a^N \bm{\mu}_a/N$, so we can rule out any long range 
orientation of the dipoles.
In all the states considered below $\langle{\bf M}\rangle\approx\mathbf{0}$, 
and no crystallization or particle segregation occurs. 

Let us now discuss the mean squared deviations. For the TDF, the common
definition of the mean squared displacement (MSD) is used,

    \begin{equation}
    W_{T}(t) = \frac{1}{N}\langle\sum_a [{\bf r}_a(t)
     - {\bf r}_a(0)]^2\rangle,
    \end{equation}

\noindent from which one can obtain  a diffusion coefficient using the
well known long-time limit (Einstein's relation) $W_{T}(t) = 6 D_{T}t$.

The description of rotational diffusion is slightly more
complex. Recall that in our case, the ODF are described by the set
$\{\hat{\bm{\mu}}_1(t),\hat{\bm{\mu}}_2(t),...,
\hat{\bm{\mu}}_N(t)\}\in \mathbb{U}^3$, where each of the vectors 
$\hat{\bm{\mu}}_a(t)$ describe a point over the unit sphere 
$\mathbb{U}^3$. As it is immediately apparent, the long-time 
limit of the diffusion of such points over a spherical surface is a static 
homogeneous covering, which gives a $W_R(t)$ that saturates to a constant 
after a sufficiently long time \cite{castro}.
There have been various efforts to define different ways of
measuring diffusion over a spherical surface. In particular, there is a
method that generates a vector $\Delta \vec{\phi}(t)$ tangential to
the surface and normal to the direction of displacement, and with the
magnitude of the traversed angle. One then concatenates these vectorial
increments to obtain a total displacement vector $\vec{\phi}(t)$
\cite{mazza-PRE-2007}. This measure of diffusion does not saturate and, 
for small angular increments $\Delta \phi$, gives a linear behavior in 
the long time limit, \emph{i.e.}, $W_{\bm{\phi}}(t) = 4 D_{\phi} t$.
However, since the $\Delta \vec{\phi}$ contributions are all tangential
to the surface of the sphere, their sum $\vec{\phi}$ drifts away from the
surface for any finite displacements, and thus, ends up predicting
normal diffusion even in cases where there is clear arrest and
where the dipoles cannot deviate more than a small fraction of $\pi$ from
its original orientation. We therefore stick with a measure of
rotational diffusion that saturates and identify arrest as those cases
where a rotational MSD does not reach (or takes an exceedingly long time
to reach) its saturation value. Specifically, we have considered the mean 
square deviations (see appendix \ref{rotational_msds}) 

\begin{equation}
W_{\theta}(t)\equiv
\frac{1}{N}\left\langle\sum_a[\Delta\theta_a(t)]^2\right\rangle,
\label{wteta}
\end{equation}

\noindent and 

\begin{equation}
W_{\sin{\theta}}(t)\equiv\frac{1}{N}\left\langle
\sum_a[\sin{\Delta\theta_a(t)}]^2\right\rangle \label{wsin}
\end{equation}

\noindent with $\Delta\theta_a(t)\equiv\theta_a(t)-\theta_a(0)$. 
In general, the results found for $W_\theta (t)$ and $W_{\sin\theta}(t)$
are quite similar and provide the same information. Therefore we will be 
reporting from now on only the results for $W_{\theta}(t)$. The observable 
$W_{\sin\theta}(t)$, however, is important because it allows to develop a 
closed formula to obtain the rotational diffusion coefficient from simulations 
(see Eqs. \eqref{wteta} and \eqref{wsin} of appendix \ref{rotational_msds})

With respect to the correlation functions, there is a common approach
which, for completeness, is briefly reviewed here. Details are carefully
discussed in Refs.
\cite{gray-molec-fluid-1984,schilling-PRE-1997,Elizondo-PRE-2014}.
One starts by considering a microscopical density of particles
at position ${\bf r}$ and orientation $\hat{\bm{\mu}}$, at time $t$,
defined by

    \begin{equation}
    \rho({\bf r},\hat{\bm{\mu}};t) = \frac{1}{\sqrt{N}}
    \sum_a \delta({\bf r} - {\bf r}_a(t))
    \delta(\hat{\bm{\mu}} - \hat{\bm{\mu}}_a(t)). \label{rhodrt}
    \end{equation}

\noindent Instead of constructing a generalized van Hove function with
variables ${\bf r}$, $\hat{\bm{\mu}}$ at time $t = 0$, and ${\bf r}'$, $
\hat{\bm{\mu}}'$ at time $t$, it is customary to consider first the
Fourier transform and spherical harmonics expansion of $\rho({\bf r},
\hat{\bm{\mu}};t)$ \cite{gray-molec-fluid-1984}, commonly referred to as
\emph{tensorial} density modes, and given by

     \begin{eqnarray}
    \hat  \rho_{lm} ({\bf k}, t) &=& \sqrt{4\pi} i^l \int_V d^3
    {\bf r}\,
       e^{i {\bf k} \cdot {\bf r}}
     \int d^2 \bm{\mu} \, Y_{lm}^*(\hat{\bm{\mu}}) \, \rho({\bf r},
      \hat{\bm{\mu}}, t), \nonumber \\
     &=& \sqrt{\frac{4\pi}{N}}i^l \sum_a e^{i {\bf k} \cdot {\bf r}_a(t)}
     \,
     Y^*_{lm}(\hat{\bm{\mu}}_a(t)),
     \end{eqnarray}

\noindent where $Y_{lm}(\hat{\bm{\mu}})$  are spherical harmonics and 
\textbf{k} is the scattering vector. From
here, the fundamental correlators are defined by a family of intermediate
scattering functions (ISFs) of the form
\cite{schilling-PRE-1997,Elizondo-PRE-2014}

     \begin{eqnarray}
     F_{lm, l'm'}({\bf k}, {\bf k}', t)
     &=& \left\langle\hat \rho_{lm}({\bf k}, 0) \, 
     \hat \rho^*_{l'm'}({\bf k}', t)\right\rangle
     \nonumber \\
     &=& \frac{4\pi }{N} i^{l'-l}\sum_{ab} \left\langle e^{i [{\bf k} \cdot          
     {\bf r}_{a}(0) - {\bf k}'\cdot {\bf r}_b(t)]} \, 
     Y^*_{lm}(\hat{\bm{\mu}}_a(0)) \, Y_{l'm'}(\hat{\bm{\mu}}_b(t))
     \right\rangle.\label{isfs_gen}
     \end{eqnarray}

Up to now no assumptions have been made about the fluid. However, one
knows that in an homogeneous system, a van Hove correlation function
does not depend on two vectors ${\bf r}$ and  ${\bf r}'$ separately, but
only on the displacement ${ |\bf r} - {\bf r}'|$.
As a result, the corresponding ISFs will only depend on one wavevector
${\bf k}$, instead of the two appearing, for instance, in eq.
\eqref{isfs_gen}.
Furthermore, in an isotropic system the van Hove function does not
depend on the two full sets of angles defining the orientations
$\hat{\bm{\mu}}$ and $\hat{\bm{\mu}'}$, but only on three angular
variables \cite{gray-molec-fluid-1984}.
Following the form of the anisotropic dipolar interaction, an
appealing choice is to use the projections of $\hat{\bm{\mu}}$ and $
\hat{\bm{\mu}'}$ over the line connecting the two dipoles (assuming some
orientation for this line which, at the end, results irrelevant), and the
dihedral angle between them. This is equivalent to choosing the line 
between the two particles as the ${\bf \hat z}$ axis, a prescription 
known as the $k$-frame choice \cite{wertheim, gray-molec-fluid-1984} 
(also referred to as \emph{interaction} frame).
Going again into Fourier space, the equivalent procedure is to rotate the
coordinate system so that the wavevector ${\bf k}$ points in the 
direction of the the ${\bf \hat z}$ axis.

As discussed in Ref. \cite{schilling-PRE-1997}, for an isotropic space 
and using well-known properties of spherical harmonics, this choice
imposes several restrictions on $F_{lm,l'm'}$. The most important of them
being that, for a nonzero correlator, one only needs to consider 
$m = m'$. In this way, one gets a right counting of variables, since the 
ISF now uses just three orientational indices. We are then left with 
correlators of the form

     \begin{equation}
     F_{ll'\, m}(k \hat {\bf z}, t)
     = \frac{4\pi}{N}i^{l'-l} \sum_{ab} <e^{i k [z_a(0) - z_b(t)]} \,
     Y^*_{lm}(\hat{\bm{\mu}}_a(0))
     \, Y_{l'm}(\hat{\bm{\mu}}_b(t))>.\label{isfs}
     \end{equation}

Finally, and as indicated before, in this contribution we are interested
only in the \emph{self} part of these correlation functions, that is,
the correlators of the positions and orientations of a single particle,

     \begin{equation}
     F^\text{S}_{ll'\, m}(k \hat {\bf z}, t)
     = \frac{4\pi}{N}i^{l-l'} \sum_{a} <e^{i k [z_a(0) - z_a(t)]} \,
     Y^*_{lm}
     (\hat{\bm{\mu}}_a(0))
     \, Y_{l'm}(\hat{\bm{\mu}}_a(t))>.
     \label{F_self_formula}
     \end{equation}

There are additional restrictions for $F_{ll'\, m}(k {\bf \hat z}, t)$
\cite{schilling-PRE-1997}. Two of them are worth mentioning here:
(\emph{i}) these correlation functions are zero unless $l - l'$ is even,
and (\emph{ii}), $F_{ll'\, m}$ is zero at time $t  = 0$ unless $l' = l$.

\subsection{Molecular Dynamics simulations}\label{details}

We have considered a Newtonian MD that uses the velocity-Verlet
integrator, for both translations and rotations.
Both sets of degrees of freedom are coupled to stochastic thermostats
that performs velocity re-scaling in the same spirit as the
Bussi-Donadio-Parrinello thermostat \cite{bussi-JCP-2007}.
Two thermostats are used because, in general, the evolution
timescales for translations and rotations are different, and the system
tends to show a difference in temperatures analogous to the
\emph{Hot-Solvent/Cold-Solute\/} problem that is pervasive in
other applications of MD, when some type of velocity
rescaling is used with only one thermostat \cite{cheng-JPC-1996}.

We have simulated a total of $N$ dipolar particles ($N_1 = N_2 = N/2$) 
in a cubic box of volume $V$, with periodic boundary conditions.
The effective packing fraction, $\eta= N\pi(\sigma_1^3+\sigma_2^3)/12V$,
and the scaled temperature, $T^*$, were used to explore different points 
in the $(\eta,T^*)$ plane.
Specifically, we have investigated state points along four different 
\emph{Paths}. Three of them consist in fixing the volume fraction
to the values $\eta = 0.2$ (\emph{Path 1}), $0.4$ (\emph{Path 2}) and 
$0.6$ (\emph{Path 3}), and varying the temperature in the range 
$0.2\leq T^*\leq2$. 
In addition, we also considered the isotherm $T^*=1$ and varied the 
concentration as $0.2\leq\eta\leq0.7$ (\emph{Path 4}).
In all cases, we have considered $N_1=N_2=4096$ for a total of $8192$ 
particles, a  number large enough so we have not detected any changes in 
the behavior of the fluid when halving $N$. 
For each state point investigated, $8$ different \emph{seeds} 
(realizations) of the system have been used to explore the available 
phase space and to improve statistics.
The time increment for $T^* = 1$ was set to $dt_1 = 5 \times 10^{-4}$, 
and for other temperatures  we use $dt = dt_1/\sqrt{T^*}$. 
The computing of the \emph{self} ISFs has been carried out performing, 
for each configuration considered, the sum indicated in Eq. 
(\ref{F_self_formula}) taking as the vertical direction each of the 
three coordinate axes, that is, using the original coordinates and then 
performing the two cyclic rotations $xyz \to zxy \to yzx$. These 
correlators are real when we consider both possible orientations for 
the ${\bf z}$ axis.
We start from a random, non-overlapping configuration, and then run a 
transient long enough to render stable values for the pressure and the
potential energy of the system

\section{Simulation results}\label{simresults}

We now present the results of the described simulations, regarding the 
dynamical properties associated with the translational and rotational 
degrees of freedom.

\subsection{Translational dynamics}\label{tdfs}

\begin{figure}
\center
{\includegraphics[width=2.6in, height=2.6in]{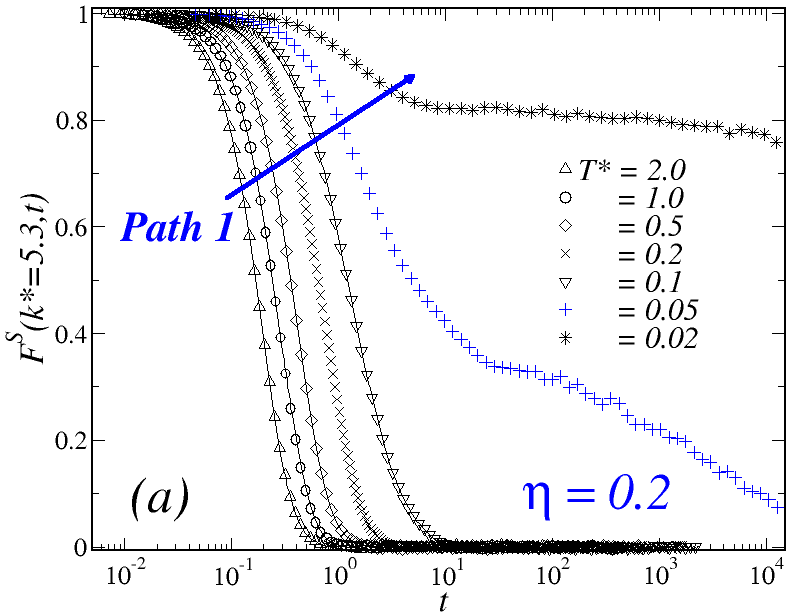}
\includegraphics[width=2.6in, height=2.6in]{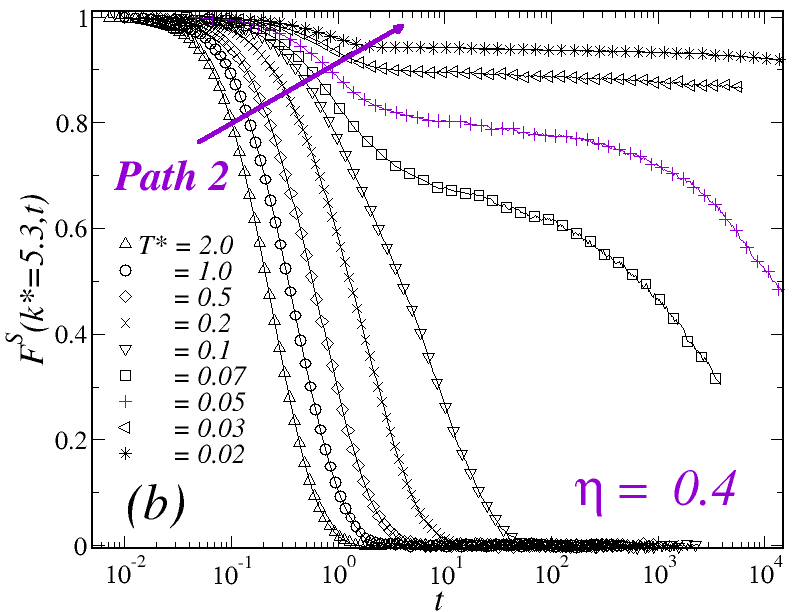}\\
\includegraphics[width=2.6in, height=2.6in]{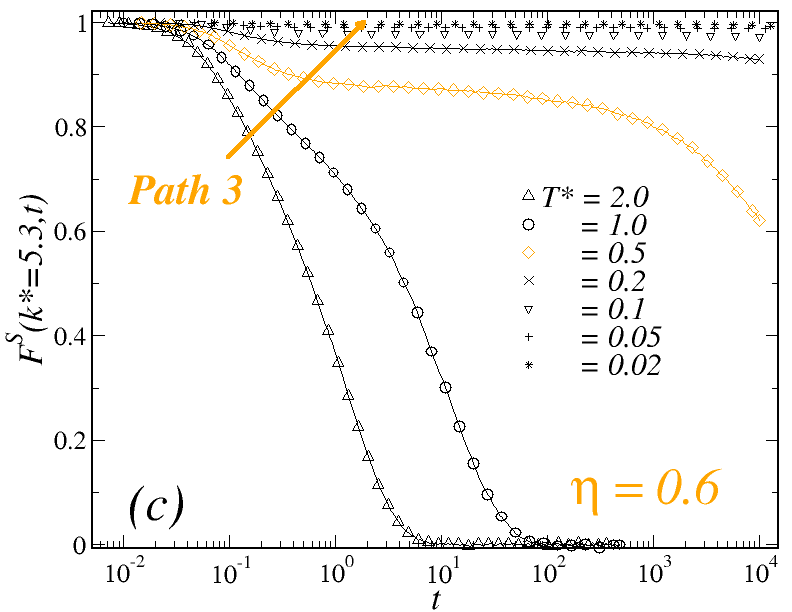}}
\includegraphics[width=2.6in, height=2.6in]{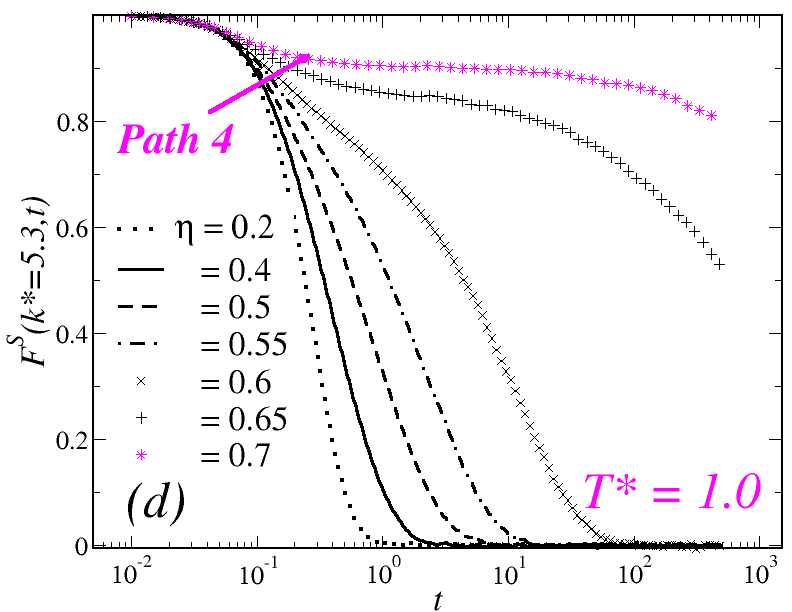}
\caption{(Color online) Time evolution of the positional-density 
\emph{self} correlation function, $F^S(k^*=5.3,t)$, with 
$k^*\equiv k\sigma_{ave}$, at three different concentrations: 
\emph{(a)} $\eta=0.2$, \emph{(b)} $\eta=0.4$, and \emph{(c)} $\eta=0.6$, 
for different temperatures (as indicated); and at fixed temperature 
\emph{(d)} $T^*=1.0$ and varying the concentration (as indicated). 
From left to right, the solid arrows intersect the curves in order of 
decreasing temperature (\emph{a-c}) or increasing density (\emph{d}).}
\label{Fig1}
\end{figure}

To study the dynamics of the TDF of the model we considered the
purely translational \emph{self}-ISF, $F_{000}^S(k,t)\equiv F^S(k,t)$, 
and the MSD, $W_T(t)$. 
As indicated above, both observables were investigated at different state 
points of the $(\eta,T^*)$ parameters space, organized for clarity in 
four \emph{Paths} that approach dynamical arrest in different ways.
In Fig.\ref{Fig1}, we summarize the results obtained for $F^S(k^*=5.3,t)$,
with $k^*\equiv k\sigma_{ave}$, at the three isochores corresponding to
\emph{(a)} $\eta=0.2$, \emph{(b)} $\eta=0.4$ and \emph{(c)} $\eta=0.6$,
and along the isotherm \emph{(d)} $T^*=1$.

At the lowest density and high temperatures considered, the system's 
behavior resembles that of a dilute HS gas (see also appendix 
\ref{appendix_gofr}). Lowering $T^*$ at fixed $\eta$ 
(\emph{i.e.} following the solid arrow of Fig.\ref{Fig1} \emph{(a)} 
from left to right, \emph{Path 1}), a gradual slowing down in the 
relaxation dynamics is observed. Specifically at $T^*=0.05$, 
$F^S$ shows an inflection point and a stretched relaxation which remains 
finite over the time-scale of the simulations. Slightly below, at $T^*=0.02$, 
the arrest of the dynamics of the TDF becomes more noticeable with the 
emergence of a high plateau, which remains practically constant up to three 
decades in time.
  
Upon increasing the concentration to $\eta=0.4$ (Fig.\ref{Fig1} \emph{(b)}, 
\emph{Path 2}) one finds similar features. 
Within the range $0.1\leq T^*\leq2$, $F^S$ becomes only moderately slower 
with respect to the previous case at comparable temperatures. 
At lower values, the \emph{self}-ISF develops a second decay mode and the 
relaxation-time increases dramatically with small variations in $T^*$. 
A transient plateau appears at $T^*=0.05$, where this 
correlation function barely changes along three decades and remains finite 
at the longest time of the simulation. The height of this plateau, 
however, is noticeably larger in comparison to that found along the 
isochore $\eta=0.2$ at the same temperature. This suggests that the state 
point ($\eta=0.4,T^*=0.05$) is closer to conditions of dynamical 
arrest than ($\eta=0.2,T^*=0.05$). 
In addition, the appearance of a higher plateaus is typically associated 
to the formation of a mechanically stronger glassy state.  
It is worth stressing that, despite the small differences in the dynamics
of the TDF along \emph{Paths 1} and $2$, a concomitant change in the 
structure of the system is also observed upon increasing $\eta$ from $0.2$
to $0.4$. These effects can be represented, for instance, by the evolution 
of the radial distribution function $g(r)$ along both isochores (see 
appendix \ref{appendix_gofr}). 
In this work, however, we specifically focus in the analysis of the slow 
dynamics towards the GT.

For the isochore $\eta=0.6$ (Fig.\ref{Fig1} \emph{(c)}, \emph{Path 3}), 
$F^S$ becomes arrested and develops a high plateau at $T^*=0.5$ (and 
below). This value in temperature is relatively larger in comparison to 
those found along the two previous cases (roughly, one order of magnitude) 
and accounts for the strong influence of crowding in the dynamical arrest 
of the translational dynamics.
These effects can be also emphasized by considering the results along the 
isotherm $T^*=1$ (Fig.\ref{Fig1} \emph{(d)}, \emph{Path 4}), where the 
features of a HS (\emph{i.e.,} density-driven) GT are displayed. In this 
case, the slowing down of the relaxation dynamics is observed for 
concentrations  $\eta\geq0.6$, where $F^S$ develops a two-steps relaxation 
pattern and eventually a non-decaying plateau at $\eta=0.7$.
Therefore, the results along \emph{Paths 3} and $4$ support the 
interpretation that the critical temperature for the arrest of the 
translational dynamics represents a rapidly increasing function of 
$\eta$ at high densities.

\begin{figure}
\center
{\includegraphics[width=2.5in, height=2.5in]{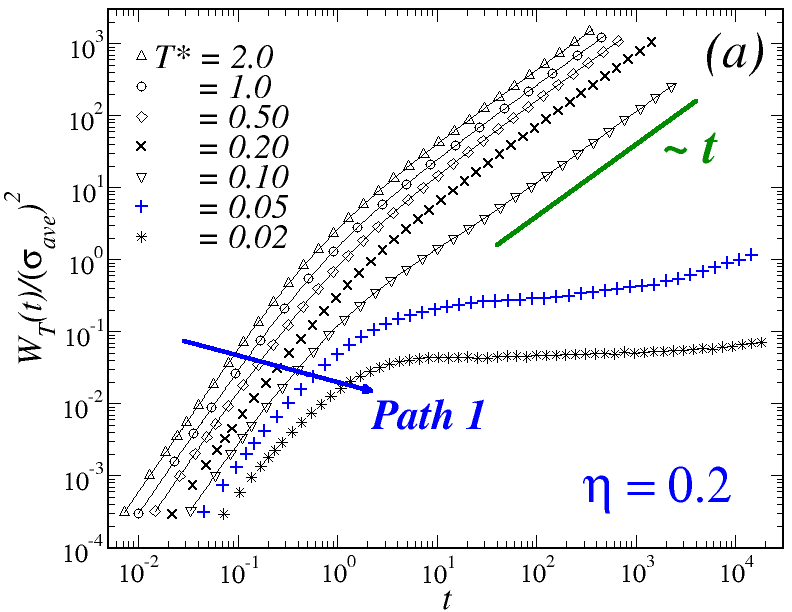}
\includegraphics[width=2.5in, height=2.5in]{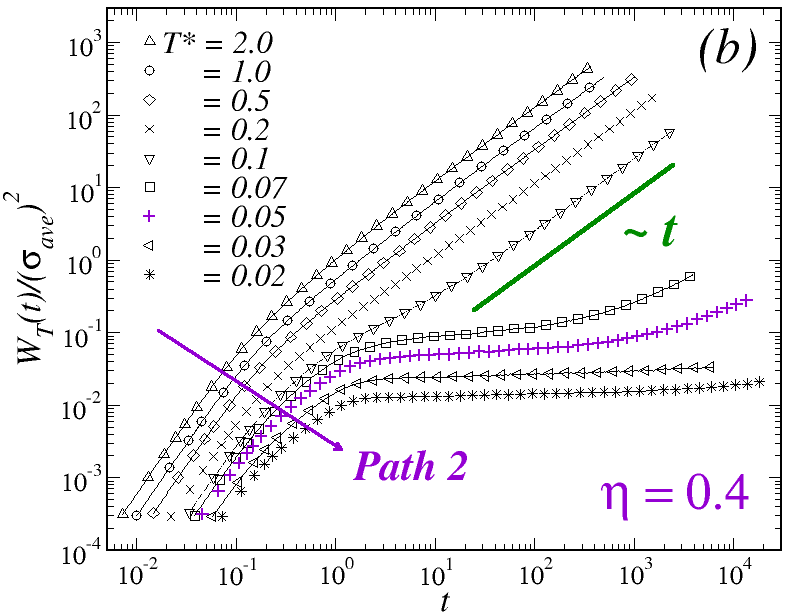}\\
\includegraphics[width=2.5in, height=2.5in]{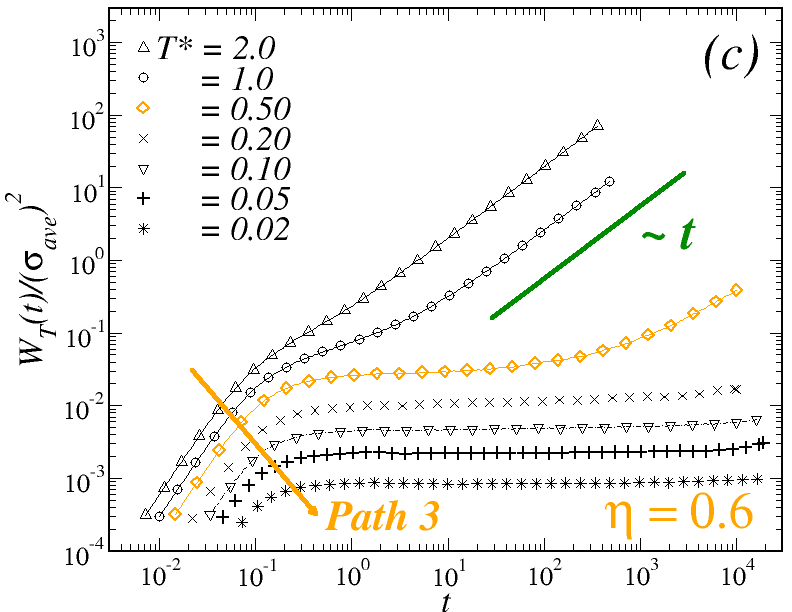}
\includegraphics[width=2.5in, height=2.5in]{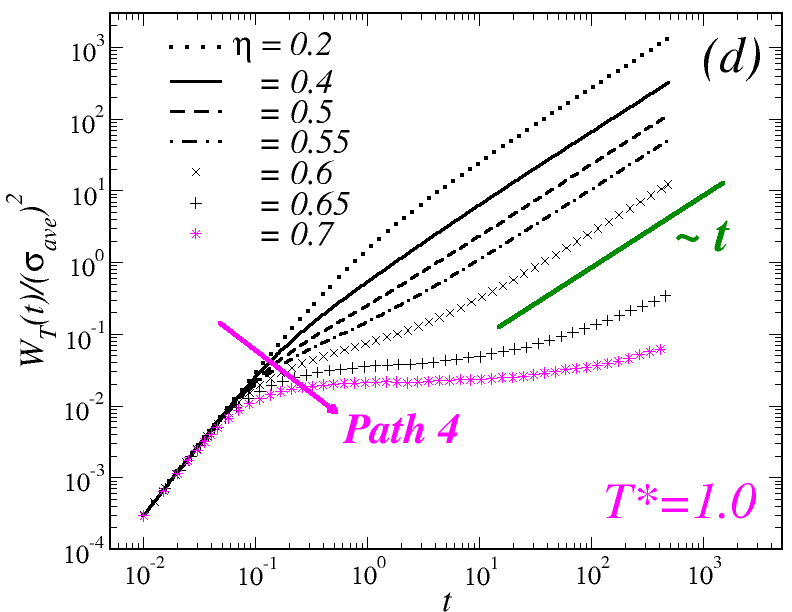}}
\caption{(Color online) Mean square displacements, $W^*(t)$ (in units of
$\sigma_{ave}^2$), corresponding  to  the isochores:
\emph{(a)} $\eta=0.2$,  \emph{(b)} $\eta=0.4$, and \emph{(c)} $\eta=0.6$
and the same temperatures as in Fig.\ref{Fig1}; 
and for the isotherm \emph{(d)} $T^*=1.0$. From left to right, the
solid black arrows intersect the curves in order of decreasing 
temperature or increasing density. The (green) solid lines are a guide
to the eye.} \label{Fig2}
\end{figure} 
 
Besides the information provided by the \emph{self}-ISF, one can also 
consider the evolution of the MSD, $W_T(t)$, along \emph{Paths 1-4}. 
Both quantities are connected in the low-wave-vector, but the MSD 
represents an easily interpreted observable quantifying particle mobility 
and long-ranged transport.  
The corresponding data are displayed in Fig. \ref{Fig2} and consistently 
describe the physical scenario outlined by $F^S$. Along the isochores 
$\eta=0.2$ and $\eta=0.4$ (Figs. \ref{Fig2}\emph{(a)} and 
\ref{Fig2} \emph{(b)}, respectively), $W_T(t)$ undergoes the typical 
evolution from ballistic ($\sim t^2$) to diffusive ($\sim t$) behavior 
for $0.1\leq T^*\leq2$. 
At lower temperatures, the particles cease to diffuse and the MSD develops 
transient plateaus followed by the emergence of a small subdiffusive regime 
for longer times. In both \emph{Paths}, this is observed to occur at 
$T^*\approx0.05$, and flat plateaus extending the observation time of the 
simulation appear for lower temperature. 

As it might be expected from the results for $F^S$ along \emph{Path 3}, 
the MSD shows arrest at $T^*=0.5$ for the isochore $\eta=0.6$ 
(Fig. \ref{Fig2}\emph{(c)}). 
Similarly, in the case of the isotherm $T^*=1$, $W_T(t)$ shows a 
slowing down with increasing density and develops 
plateaus for $\eta\geq0.65$. This confirms that $F^S$ and $W(t)$
provide the same physical scenario regarding the dynamical arrest 
of the TDF.

To estimate the maximum possible displacement of the particles approaching 
to conditions of dynamical arrest, one can use the long-time value of the MSD. 
In the case of dense systems described by steep repulsive potentials, 
the square-root of this value is commonly referred to as the localization 
length, $l$, and represents a measurement of the local confinement, 
since it provides the displacement inside nearest neighbors \emph{cages}. 
The quantity $l(\eta,T^*)\equiv\sqrt{W_T(\eta,T^*;t^*=10^2)}$, for instance, 
shows that the particles only become slightly more localized upon increasing
$\eta$ from $0.2$ to $0.4$, despite the dynamical (and structural) differences 
found along \emph{Paths 1} and $2$ (\emph{i.e.,} 
$l(\eta=0.2,T^*=0.05)\approx l(\eta=0.4,T^*=0.05)$).
In addition, both $l(\eta=0.6,T^*=0.5)$ and $l(\eta=0.7,T^*=1)$ provide
essentially the same value, $l\approx0.1\sigma_{ave}$,  which 
corresponds to a \emph{cage size} of approximately $10\%$ of the 
characteristic diameter, a typical feature of a HS glass. 

In summary, the results for $F^S(k^*,t)$ and $W_T(t)$ outline the main 
features of arrested dynamics of the TDF of the simulated model. 
At low and intermediate concentrations, the dynamics freezes at nearly 
the same temperature and shows essentially the same localization length 
for the constitutive particles, but with some differences in the 
corresponding relaxation times, decay patterns, and also at the structural 
level. The results for the high density regime, instead, show that the 
critical temperature for dynamical arrest behaves as a monotonically and 
rapidly increasing function of the concentration, and that the dynamics of 
the TDF undergo a GT in the fashion of a HS system.

\subsection{Orientational dynamics}\label{odfs}

We now turn our discussion to the dynamics of the ODF.
For this, we have considered a set of orientationally sensitive 
correlation functions, $F^S_{ll'm}(k,t)$, with $l=l'>0$ and 
defined by Eq. \eqref{F_self_formula}; and the angular mean squared
deviations, $W_{\theta}(t)$ and $W_{\sin\theta}(t)$, given by
Eqs. \eqref{wteta} and \eqref{wsin}. For simplicity, we will restrict ourselves
to the analysis of ISFs of rank:  $1\leq l=l'\leq2$. Furthermore, 
we have found that both $F_{221}^S$ and $F_{222}^S$ provide 
essentially the same information as $F_{220}^S$.
Therefore, we specifically focus on the analysis of the correlation 
functions $F_{110}^S(k,t)$, $F_{111}^S(k,t)$, and $F_{220}^S(k,t)$. 
As mentioned before, a similar behavior is encountered for 
$W_{\theta}(t)$ and $W_{\sin \theta}(t)$, so we only report the 
former.

\begin{figure*}
\center
{\includegraphics[width=2.1in, height=2.1in]{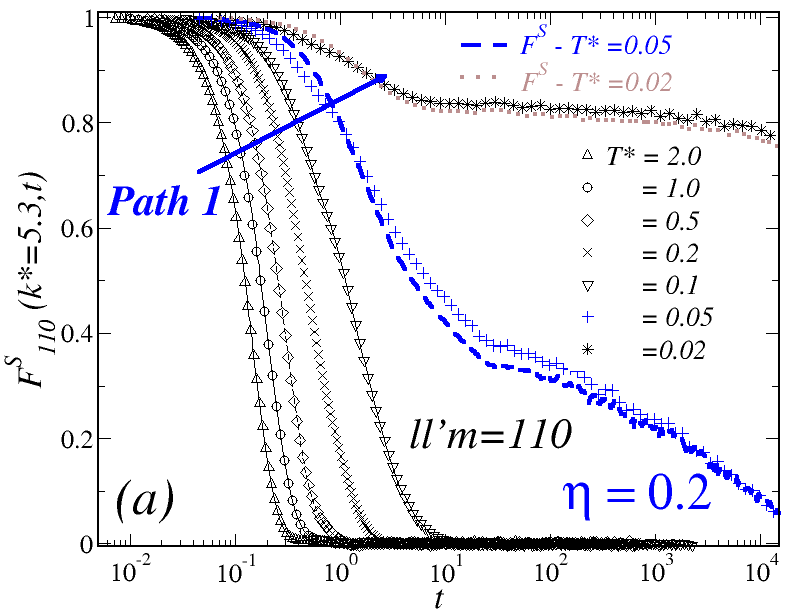}
\includegraphics[width=2.1in, height=2.1in]{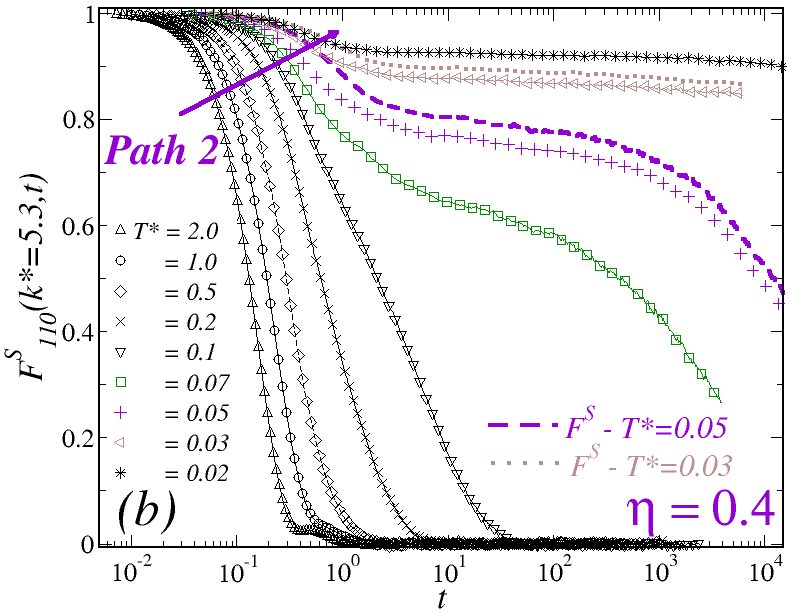}
\includegraphics[width=2.1in, height=2.1in]{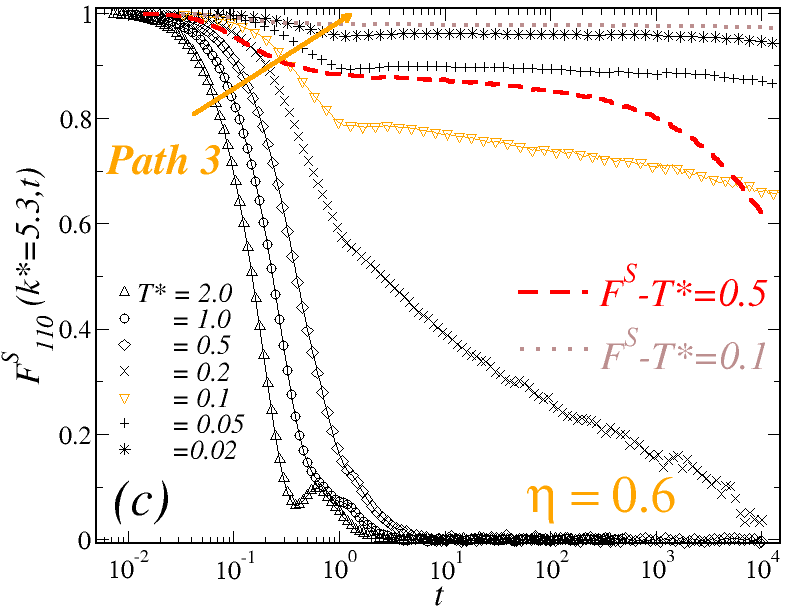}\\
\includegraphics[width=2.1in, height=2.1in]{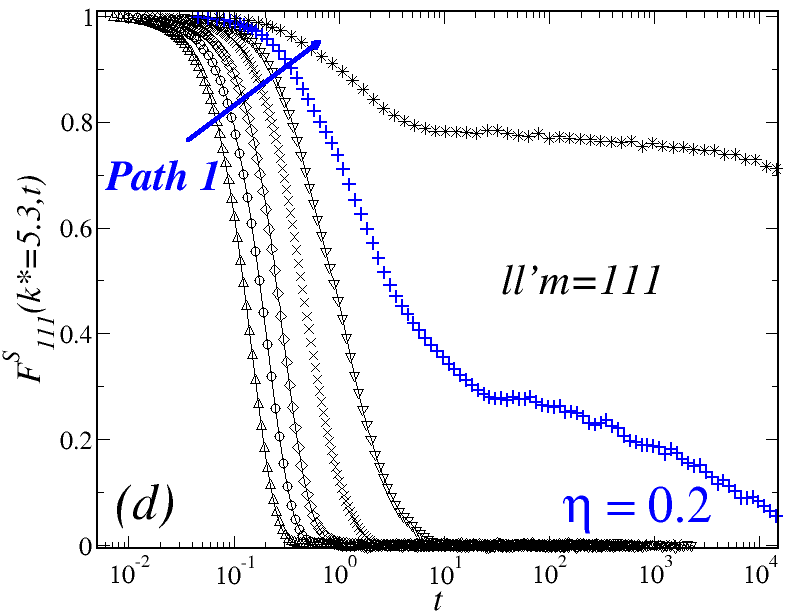}
\includegraphics[width=2.1in, height=2.1in]{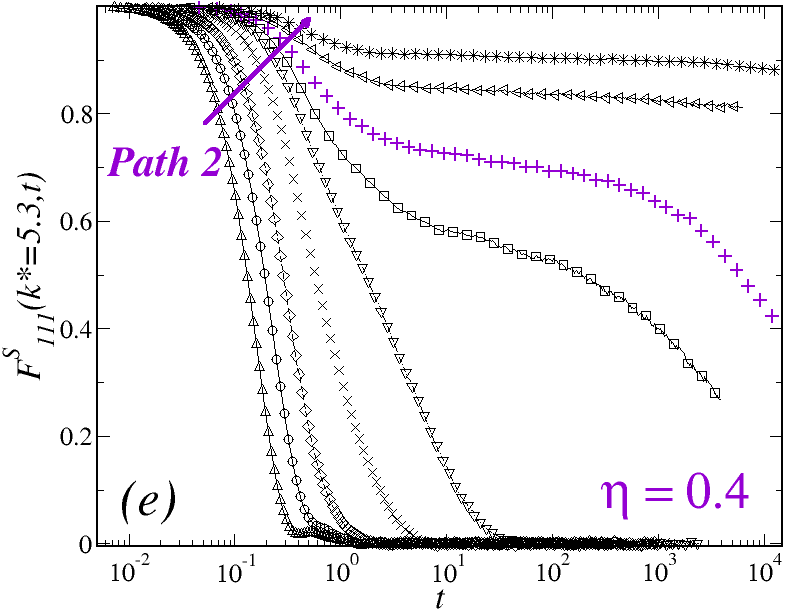}
\includegraphics[width=2.1in, height=2.1in]{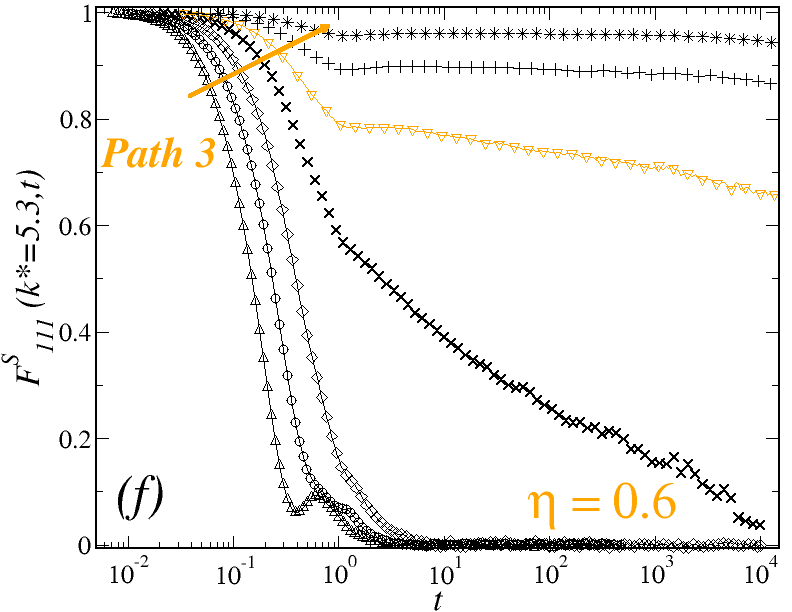}\\
\includegraphics[width=2.1in, height=2.1in]{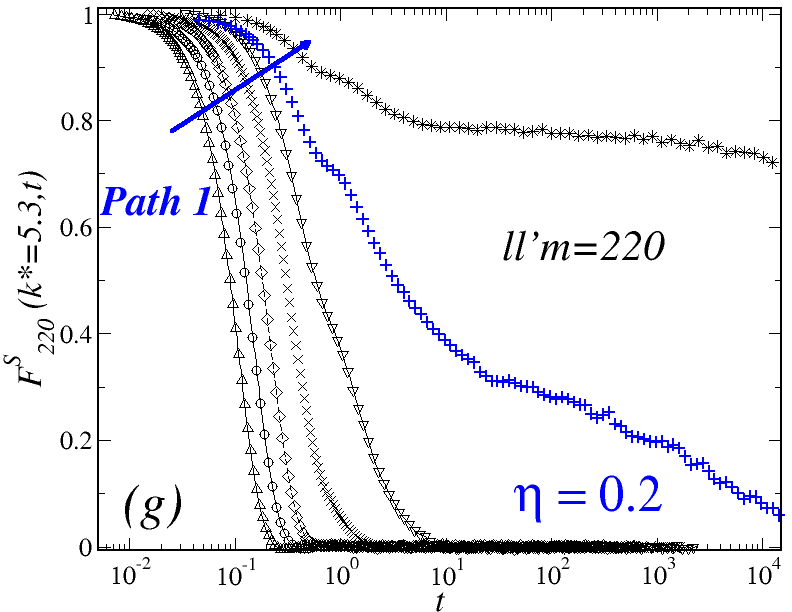}
\includegraphics[width=2.1in, height=2.1in]{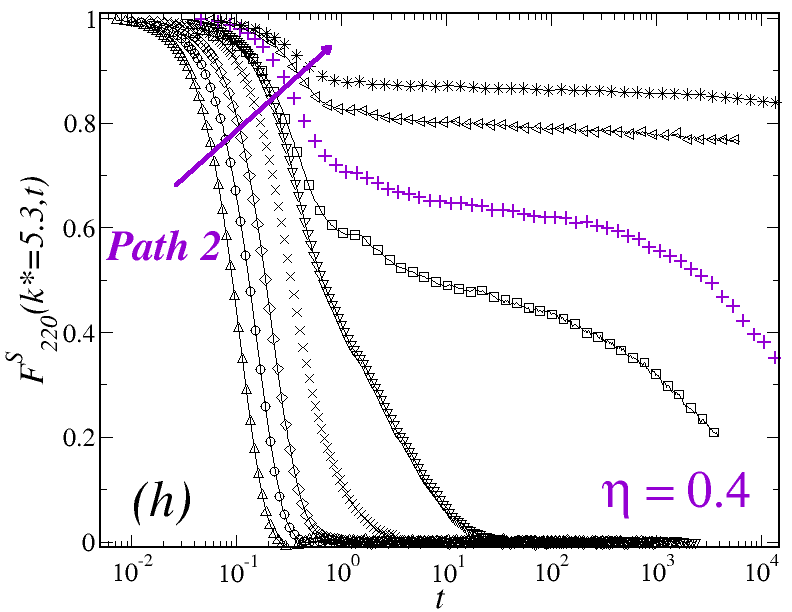}
\includegraphics[width=2.1in, height=2.1in]{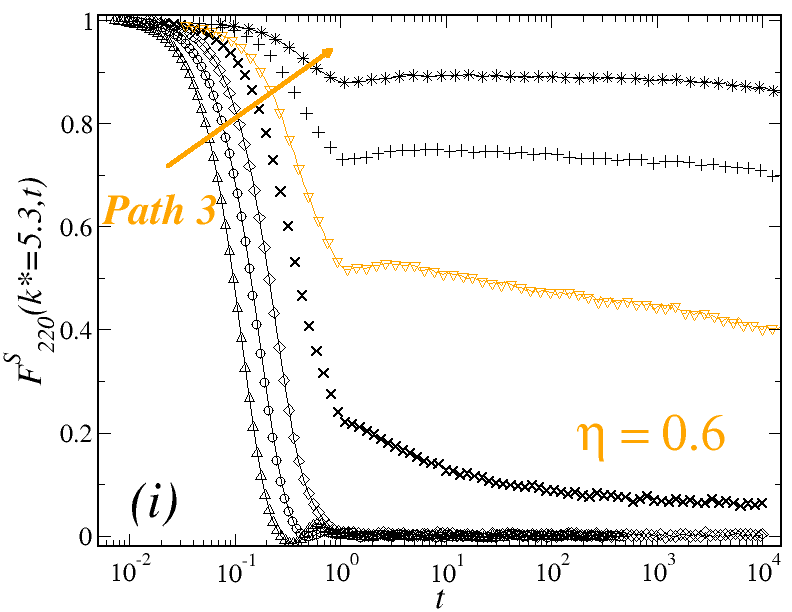}}
\caption{(Color online) Time evolution of the correlation functions
$F_{110}(k^*,t), F_{111}(k^*,t)$ and $F_{220}(k^*,t)$, evaluated at
$k^*=5.3$, at three different concentrations $\eta=0.2$ \emph{(a)-(d)-(g)},
$\eta=0.4$ \emph{(b)-(e)-(h)}, $\eta=0.6$ \emph{(c)-(f)-(i)}, and at
different temperatures (as indicated). For reference, the dashed and 
dotted lines in \emph{(a)}-\emph{(b)}-\emph{(c)} illustrate the arrested 
behavior of $F^S(k,t)$ previously shown in Fig.\ref{Fig1}.} \label{Fig3}
\end{figure*}

The results for the orientational ISFs along the three isochores 
previously considered (\emph{Paths 1,2,3}), evaluated at $k^*=7$, 
are summarized in Fig.\ref{Fig3}. Either at low (first column, 
Figs. \ref{Fig3}\emph{(a)}-\emph{(d)}-\emph{(g)}) or intermediate 
concentration (middle column, Figs. 
\ref{Fig3}\emph{(b)}-\emph{(e)}-\emph{(h)}), 
$F_{110}^S$, $F_{111}^S$, and $F_{220}^S$ display essentially
the same relaxation patterns and become gradually slower with 
decreasing temperature. Remarkably, all the correlation functions also  
follow the same trends of $F^S$ along \emph{Paths 1} and $2$, including 
their eventual arrest at $T^*=0.05$ (for clarity, the arrested behavior 
of $F^S$ is displayed again in Figs. \ref{Fig3}\emph{(a)-(b)-(c)} as 
the dashed and dotted lines).
These features indicate a strong coupling in the dynamics of the TDF 
and ODF, thus evidencing their simultaneous arrest upon cooling.

This is to be contrasted with the scenario observed for $\eta=0.6$
(right column in Fig. \ref{Fig3}). First, let us notice that the 
orientational ISFs remain ergodic (\emph{i.e.,} decay to zero) at 
$T^*=0.5$ (open diamonds in Figs.\ref{Fig3} \emph{(c)}-\emph{(f)}-
\emph{(i)}), whilst $F^S$ undergoes dynamical arrest 
(red dashed line of Fig.\ref{Fig3}\emph{(c)}). 
Lowering the temperature to $T^*=0.2$, $F^S_{110}$ and $F^S_{111}$ 
develop transient plateaus, which seem to decay slowly to zero in the 
time scale of the simulations, whereas $F^S_{220}$  develops a slightly 
different relaxation pattern characterized by a fast initial decay and 
followed by a stretched relaxation. These features indicate the 
existence of a transition to \emph{partially} arrested states for the 
fluid, where the dynamics of the TDF undergoes a GT whereas the ODF 
remain ergodic. 
Furthermore, at $T^*=0.1$ the three correlation functions also become 
arrested, indicating another type of transition where the ODF may 
undergo a GT starting from a \emph{partially} arrested state. 
Notice that this transition for the ODF occurs at a slightly higher 
temperature with respect to \emph{Paths 1} and $2$.

\begin{figure*}
\center
{\includegraphics[width=2.1in, height=2.1in]{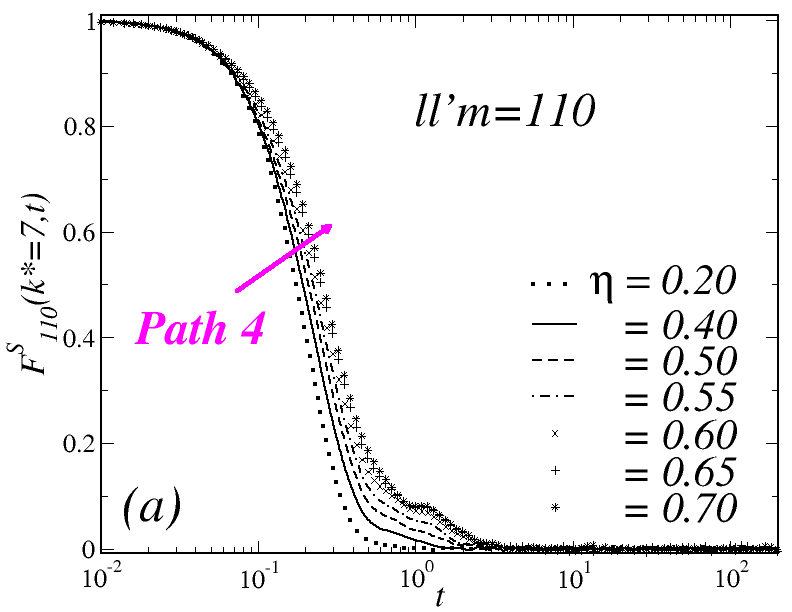}
\includegraphics[width=2.1in, height=2.1in]{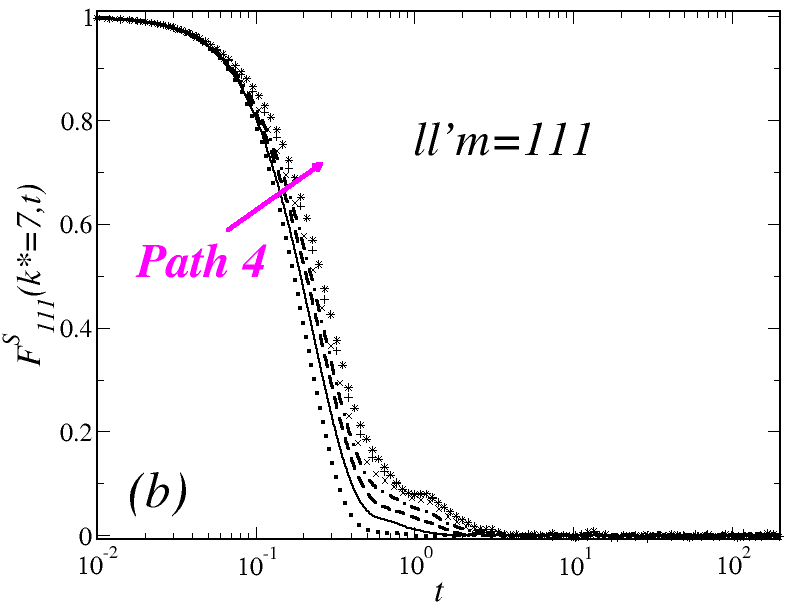}
\includegraphics[width=2.1in, height=2.1in]{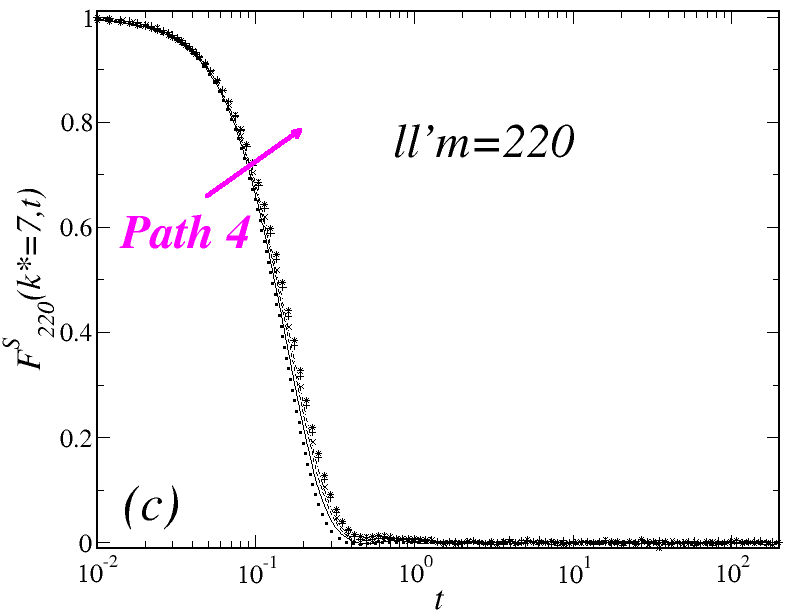}}
\caption{(Color online) Time evolution of the correlation functions
\emph{(a)} $F_{110}(k^*,t)$, \emph{(b)} $F_{111}(k^*,t)$ and
\emph{(c)} $F_{220}(k^*,t)$ evaluated at $k^*=5.3$, along the isotherm 
$T^*=1$, for different concentrations (as indicated).} 
\label{isoterm1}
\end{figure*}

The existence of \emph{partially} arrested states is also illustrated 
by the results shown in Fig. \ref{isoterm1} for the orientational ISFs 
along the isotherm $T^*=1$, where the three correlation functions remain 
ergodic and practically unaffected by increasing the concentration up to 
$\eta=0.7$. This is in clear contrast with the behavior of $F^S$, 
which hardly relaxes to zero for $\eta=0.65$ and becomes arrested at 
$\eta\geq0.7$, as already shown in Fig. \ref{Fig1} \emph{(d)}.

\begin{figure}
\center
{\includegraphics[width=2.5in, height=2.5in]{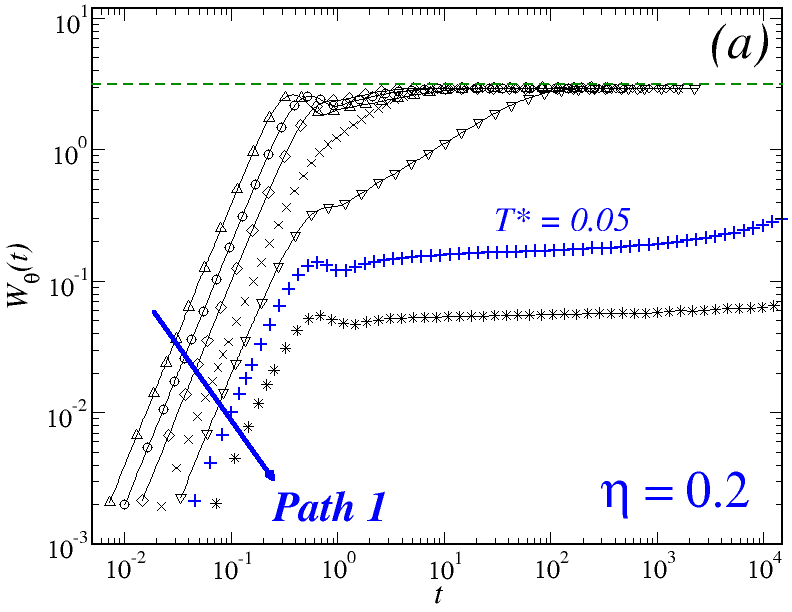}
\includegraphics[width=2.5in, height=2.5in]{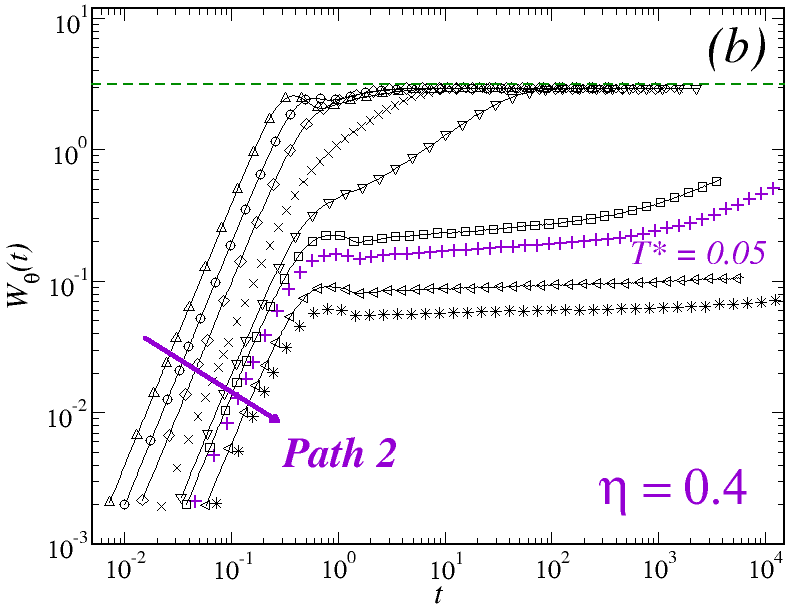}\\
\includegraphics[width=2.5in, height=2.5in]{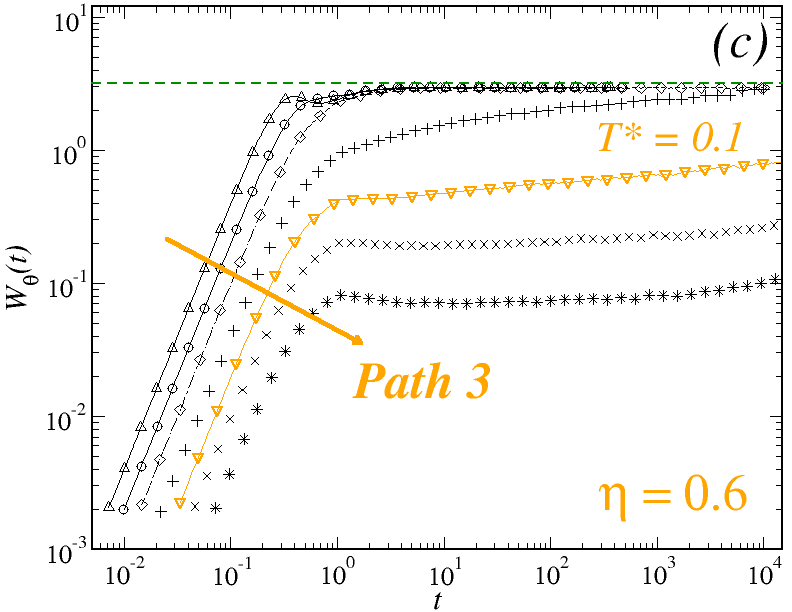}
\includegraphics[width=2.5in, height=2.5in]{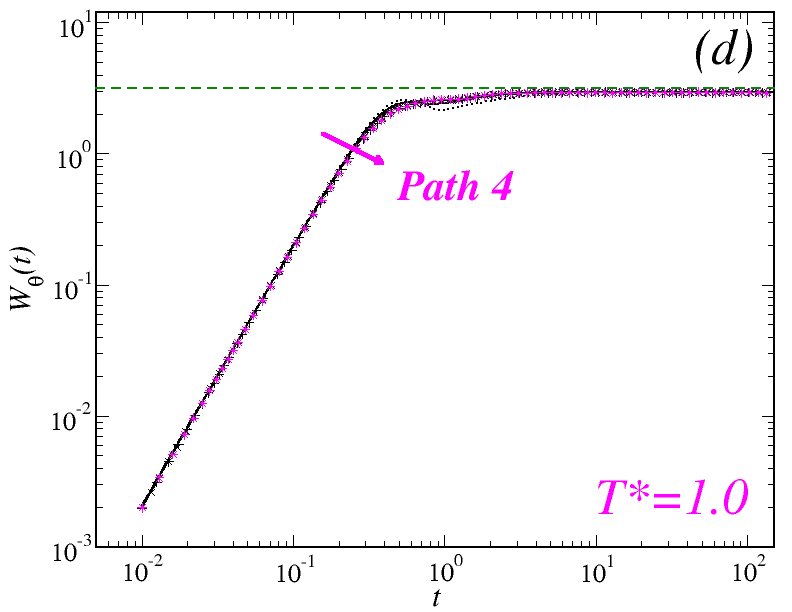}}
\caption{(Color online) Behavior of the angular mean square displacement, 
$W_{\theta}(t)$, at the three isochores \emph{(a)} $\eta=0.2$, 
\emph{(b)} $\eta=0.4$ and \emph{(c)} $\eta=0.6$, for different 
temperatures; and for the isotherm (d) $T^*=1.0.$ Symbols
are used exactly as in Figs. \ref{Fig3} and \ref{Fig4} above to represent 
the state points. The (green) dashed horizontal lines indicate the ergodic 
long time value of $W_{\theta}$ and are used as guide to the eye.} 
\label{Fig4}
\end{figure}

The angular MSD, $W_{\theta}(t)$,  describes consistently the same
scenario for the arrest of the orientational dynamics. 
Along \emph{Paths 1} and $2$ (Fig. \ref{Fig4}\emph{(a)} and 
Fig. \ref{Fig4}\emph{(b)}, respectively) $W_{\theta}(t)$ reaches its 
ergodic saturation value within the range $0.1\leq T^*\leq2$. 
As it may be expected, the time elapsed to reach this limit becomes 
progressively larger with decreasing temperature. 
Below this range, $W_{\theta}(t)$ saturates to a 
smaller value, thus indicating arrest in the diffusion of the ODF. 
For the isochore $\eta=0.6$, the angular MSD rapidly reaches the 
ergodic value for temperatures down to $T^*=0.5$. At $T^*=0.2$, however, 
this requires the whole time-window of the simulations. For even 
lower temperatures, within the same time window,
$W_{\theta}(t)$ appears fully arrested. Finally, this observable remains 
unaffected at the isochore $T^*=1.0$ (Fig. \ref{Fig4}\emph{(d)})

To summarize, the MD simulations exhibit three different scenarios for 
dynamical arrest in the model. The first occurs at low and intermediate concentrations 
upon lowering the temperature and is mainly characterized by the 
simultaneous arrest of the TDF and ODF, with both dynamics being strongly 
coupled. Neither the critical temperature of this transition, nor the corresponding localization length of the 
constitutive particles,  vary significantly with $\eta$, despite qualitative changes observed in the structural 
behavior of the system upon increasing the density. 
A second scenario is observed at high densities and high-to-intermediate 
temperatures, where only the dynamics of the TDF undergoes arrest whilst 
the ODF remain ergodic. This indicates the possibility of partially 
arrested states to occur, either through an isochorical drop in temperature, 
or an isothermal compression. This transition presents the features of the 
GT in dense HS fluids. 
Finally, a third possibility involves the arrest of the orientational 
dynamics by cooling down the system from a \emph{partially} arrested 
state. This was observed to occur at temperatures only slightly higher 
with respect to those describing simultaneous arrest.

\section{Theoretical development}\label{Theoryanddiagrams}

To provide a more comprehensive picture of the dynamical arrest 
transitions of a dipolar fluid, we now consider the SCGLE theory. 
Specifically, we use this theoretical framework to obtain the GT 
lines in the full parameter space of the system, leading to the 
development of an arrested states diagram which identifies the regions 
in the $(\eta,T^*)$-plane where the system remains ergodic, the regions 
where it becomes \emph{partially} or fully arrested, and the boundaries 
between such regions. This kinetic arrest diagram complements the usual 
\emph{equilibrium} phase diagram \cite{goyal1,goyal2,goyal3}. 
As we show below, the physical scenario outlined by the SCGLE describes 
consistently the dynamical features of the simulations and organizes 
qualitatively the different state points and \emph{Paths} studied with 
MD, thus providing mutual support between the results of the two 
independent approaches.

The use of the SCGLE for this purpose is particularly helpful because 
the precise determination of GT points from simulations is notoriously 
difficult. Close to a transition, one typically encounters strong 
instabilities and very slow dynamics with history dependence, 
which renders the required computational time excessively large.
Furthermore, the common protocols to estimate GT points in simulations 
are normally based on extrapolations of either the divergence of the 
relaxation time of the ISFs or the diffusion coefficients, and are 
therefore prone to errors, since this intrinsically involves a large 
uncertainty in the choice of the specific extrapolation function and 
the fit range.

\subsection{Arrested states diagram}

We briefly recall some technical aspects regarding the determination 
of GT lines within SCGLE (for more details the reader is referred to 
appendix \ref{theory} below, and to Ref. \cite{Elizondo-PRE-2014}).
The theory provides a self-consistent system of equations describing 
the full wave-vector and time dependence of the \emph{diagonal}
($l=l'$) ISFs, $F_{ll'm}(k,t)\delta_{ll'}\equiv F_{lm}(k,t)$, 
and their corresponding \emph{self} counterparts 
$F^S_{ll'm}(k,t)\delta_{ll'}\equiv F^S_{lm}(k,t)$.
Close to conditions of dynamical arrest, one can develop a generic 
asymptotic solution for these equations, leading to a new closed 
set of equations for the non-decaying $k$-components of both
$F_{lm}(k,t)$ and $F^S_{lm}(k,t)$, commonly 
referred to as non-ergodicity parameters (NEP), and defined as 
$\displaystyle{\lim_{t\to\infty}F_{lm}(k,t)}/F_{lm}(k,t=0)\equiv 
f_{lm}(k)$ and
$\displaystyle{\lim_{t\to\infty}F^S_{lm}(k,t)}\equiv f^S_{lm}(k)$.

Upon a well defined set of approximations, the equations for $f_{lm}(k)$
and $f^S_{lm}(k)$ can be rewritten as only two equations for the parameters 
$\gamma_T$ and $\gamma_R$, which play the role of order parameters in the 
determination of the ergodic-to-non-ergodic transitions of the dynamics of, 
respectively, the TDF and ODF (Eqs. \eqref{gat} and \eqref{gar} of 
appendix \ref{theory}) . 
In terms of these quantities, fully ergodic (\emph{fluid}) states are 
described by the condition that both, $\gamma_T^{-1}$ and $\gamma_R^{-1}$, 
are equal to zero. Any other solution indicates loss of ergodicty 
(\emph{i.e.,} dynamical arrest) in one or both degrees of freedom. For 
instance, the condition $\gamma_T^{-1}\neq0$ and $\gamma_R^{-1}=0$ 
describes  partially arrested states, where only the TDF undergo a GT, 
whereas the dynamics of the ODF remains ergodic. Similarly, the condition 
$\gamma_T^{-1}\neq0$ and $\gamma_R^{-1} \neq0$ describes arrest in both 
degrees of freedom, \emph{i.e.,} fully arrested states. 

To solve the equations for $\gamma_T$ and $\gamma_R$, one requires the 
previous determination of the diagonal elements, $S_{lm}(k)$, of 
the spherical harmonics expansion of the static structure factor (SSF)
$S(k,\bm{\mu},\bm{\mu}')$, at each state point of the parameters space 
(notice that $S_{ll'm}(k)=F_{ll'm}(k,t=0)$ and 
$S_{ll'm}(k)\delta_{ll'}\equiv S_{lm}(k)$). In general, this might pose
a non-trivial problem on its own right. To simplify the theoretical 
calculations as much as possible, we have approximated 
the $S_{lm}(k)$ components of the simulated model system by a 
softened-core version of Wertheim's solution for the mean 
spherical model (MSM) of a dipolar hard-sphere fluid 
(DHS) \cite{wertheim}. The specific details of this approximation are 
provided in appendix \ref{append_msa}.

\begin{figure}
\center
{\includegraphics[width=4.2in, height=4.2in]{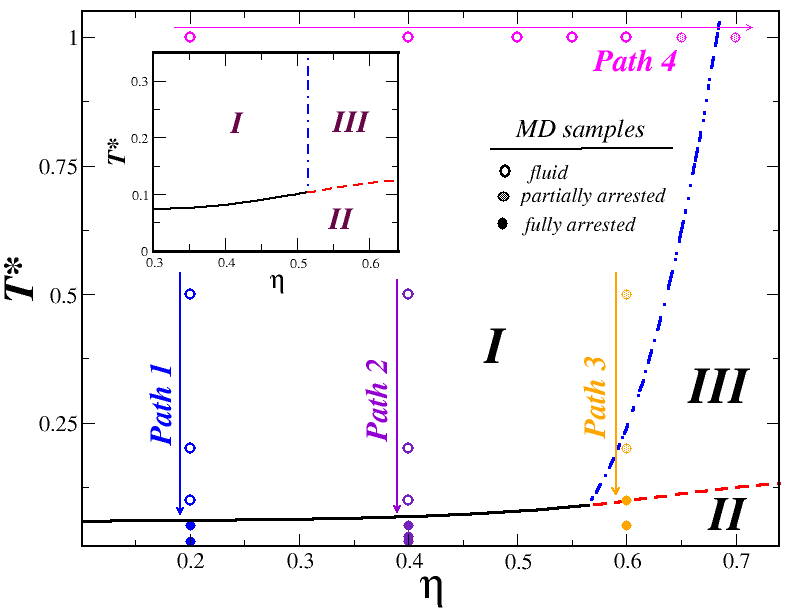}}
\caption{State space of the dipolar fluid. The different symbols
and arrows denote, respectively, the state points and \emph{Paths} 
studied via MD simulations. Open circles are used to represent the
ergodic samples, half-filled circles describe samples where 
partially arrested states are observed, solid circles 
account for full arrest. The lines delimiting regions \textbf{\emph{I}}, 
\textbf{\emph{II}} and \textbf{\emph{III}} are predictions of the
SCGLE for three distinct GT lines using the approximation described
in appendix \ref{append_msa} for the static structure factor projections, 
$S_{lm}(k)$. Inset: analogous results of the SCGLE for a DHS system
previously reported in Ref. \cite{Elizondo-PRE-2014} .} 
\label{diagram}
\end{figure}

Hence, one uses $S_{lm}(k)$ to solve the equations for $\gamma_T$ 
and $\gamma_R$ at every state point of the ($\eta,T^*$) plane. This
leads to the identification of three kinetically different regions 
separated by three distinct transition lines. Our results are summarized 
in Fig. \ref{diagram}, where we also show the location of the state 
points and \emph{Paths} investigated via MD. 
Based on the theory, the $(\eta,T^*)$ space can be partitioned as 
follows: 
a \emph{fluid} region (\textbf{\emph{I}}) where the dynamics of both 
TDF and ODF remain ergodic ($\gamma_T^{-1}=\gamma_R^{-1}=0$). This region 
is delimited by two different boundaries. One of these boundaries (black 
solid line) describes the transitions to \emph{fully arrested} states 
(\textbf{\emph{II}}), where $\gamma_T$ and $\gamma_R$ become finite 
simultaneously, so both degrees of freedom are dynamically arrested. 
The second boundary (blue dashed-pointed line), instead, describes the 
transitions to \emph{partially arrested} states (\textbf{\emph{III}}) 
where only the TDF dynamics undergo arrest ($\gamma_T$-finite) whereas 
the ODF remain ergodic ($\gamma_R^{-1}=0$). 
Finally, a third line (red dashed) separates regions \textbf{\emph{II}} 
and \textbf{\emph{III}}, thus describing the arrest of the ODF under the 
condition that the TDF were previously arrested. 

The \textbf{\emph{I}}$\to$\textbf{\emph{II}} transition line represents 
the behavior of the critical temperature, $T^{*(g)}_{I\to II}(\eta)$, as 
a function of the volume fraction as predicted by the SCGLE. 
It describes a monotonically (and very slowly) increasing function of 
$\eta$ and, noticeably,  $T^{*(g)}_{I\to II}(\eta=0.2)\approx 
T^{*(g)}_{I\to II}(\eta=0.4) \approx 0.06$, which is in remarkable 
agreement with our previous findings with MD along \emph{Paths} $1$ and $2$. 

The theory predicts that this line bifurcates into two distinct lines
upon increasing the concentration ($\eta\approx 0.57$). The branch 
describing the  \textbf{\emph{I}}$\to$\textbf{\emph{III}} transition 
shows that the critical temperature $T^{*(g)}_{I\to III}(\eta)$ increases 
significantly with small increments in $\eta$. This implies strong 
effects of crowding in the slowing down of the translational dynamics and 
a remarkable decoupling between translational and orientational dynamics as 
this transition is approached. 
This feature is also qualitatively consistent with the scenario outlined 
by the simulations along \emph{Paths} $3$ and $4$, although some small
quantitative differences are observed. For example,  the theory seems to 
slightly overestimate the critical volume fraction of the bifurcation point 
and of the \textbf{\emph{I}}$\to$\textbf{\emph{III}} transition.
This may be attributed to deviations of the simplified model assumed 
in the theory from the simulated system, and also to the approximations 
carried out to determine the structure factor projections, $S_{lm}(k)$ 
(appendix \ref{append_msa}). Nevertheless, the overall physical scenario 
is essentially the same.
 
Starting from a state point inside the partially arrested region 
\textbf{\emph{III}} and decreasing the temperature one crosses 
the \textbf{\emph{II}}$\to$\textbf{\emph{III}} transition line, 
below which, the orientational dynamics of the positionally-frozen 
dipoles also becomes arrested. Hence, this line describes a scenario
similar to a \emph{spin glass}-like transition. Notice that the 
critical temperature $T^{*(g)}_{II\to III}(\eta)$ appears almost 
independent on $\eta$ and satisfies $T^{*(g)}_{II\to III}\approx0.1$, 
a value in good quantitative agreement with the simulation results for 
\emph{Path 3}.

One can also also compare these results against the predictions of the 
SCGLE corresponding to a DHS model \cite{Elizondo-PRE-2014}, \emph{i.e.,} 
a system where the short-ranged and purely repulsive contribution to the 
potential is represented by the discontinuous (athermal) HS potential, and 
not by the soft (thermal) WCA interaction considered in both theory and 
simulations for this work. This is shown in the inset of Fig. \ref{diagram}. 
Clearly, the main difference between the two diagrams only refers to the  
slope of the \textbf{\emph{I}}-\textbf{\emph{III}} transition line.  
In the case of the DHS model, this is described by a vertical line at 
$\eta^{(g)}_{DHS}\approx0.52$, which results from the athermal nature of 
the discontinuous HS potential and, consequently, of the 
\textbf{\emph{I}}-\textbf{\emph{III}} transition in the case of the 
DHS model. 
Crucially, however, the topology of the two diagrams is identical despite 
the rather different range of the anisotropic dipolar forces and torques 
in each case: the DHS model uses a genuine dipole-dipole 
potential, which considers a $r^{-3}$ contribution in all the space. In 
the simulated system, in contrast, we have truncated this interaction for 
simplicity. 
This indicates that the topology of the dynamical arrest scenario of a 
dipolar fluid is essentially determined by the anisotropy of the dipolar 
tensor 
$D(\mathbf{r}_{ab},\bm{\hat{\mu}}_a,\bm{\hat{\mu}}_b)\equiv
\bm{\hat{\mu}}_a\cdot\bm{\hat{\mu}}_b
-3(\bm{\hat{\mu}}_a\cdot\hat{\mathbf{r}}_{ab})
(\bm{\hat{\mu}}_b\cdot\hat{\mathbf{r}}_{ab})$ (see Eq. \eqref{dipdip}
above), and not by the specific range of the pair potential.  
This provides support to the systematic use of truncated anisotropic 
potentials for the study of glassy behavior in dipolar systems which,
as mentioned above, simplifies considerably the simulations, since one
avoids the specialized treatment of long-ranged forces and torques.

\section{Conclusions}\label{conclusions}

Molecular dynamics simulations and theoretical calculations were combined 
to investigate the glassy behavior and different \emph{non-equilibrium} 
phases and transitions in a dipolar fluid. The system was modeled as a 
collection of $N$ spherical particles interacting via a soft-repulsive 
potential coated with an anisotropic contribution that retains the 
directional features of the  dipole-dipole forces, but which neglects its 
long-ranged character. 
This is a reasonable approximation for dipolar colloids suspended in highly 
dielectric solvents. An advantage of this modeling procedure is that the 
simulations are not too computationally exhaustive.

We have studied the dynamics associated to the translational and 
orientational degrees of freedom, at different regions of the parameter 
space of the system, and considering distinct pathways to dynamical arrest. 
The detailed analysis of several correlation functions and of two types 
of mean square deviations along the distinct \emph{Paths}, provided evidence 
of the occurrence of three types of dynamical arrest transitions in a 
dipolar fluid.

At small and intermediate volume fractions, one observes the 
simultaneous arrest of the dynamics of both degrees of freedom on 
cooling, occurring at a critical temperature 
$T_{\emph{I}\to\emph{II}}^{(g)}\approx0.05$. 
Despite some qualitative changes observed in the structure of the 
simulated system, this transition seems to not depend crucially on 
the concentration. Both simulations and theory support this 
interpretation. 
At high concentration, instead, a bifurcation scenario for the glass 
transition is found. In this regime, a decoupling in the dynamics
of each degree of freedom leads to another type of transition describing 
partially arrested states, where only the translational motion becomes 
arrested, but with the orientations remaining ergodic. In this regard, 
it is also important to notice that neither in the 
simulations nor in the theory we observed any hint of the other possible mixed 
state, in which  the translational motion remains ergodic, but with the orientations become arrested.  
Finally, a third kind of transition can be reached starting from a partially 
arrested state and decreasing the temperature, with the orientational 
correlations displaying arrest under the condition that the translational 
dynamics was already arrested.

The physical scenario outlined by the simulations is qualitatively 
(and semi quantitatively) consistent with the results of the SCGLE 
theory. The latter, however, allows us to develop a more generic description 
of dynamical arrest in the dipolar fluid.
This was conveniently summarized in an arrested states diagram, which  
results topologically identical to that of a dipolar hard sphere fluid, 
where different short and long-range interactions are considered. 
This indicates that our results may be generic to systems with competing 
dynamics arising from dipolar anisotropic forces and torques. 
These observations could also be relevant for the understanding of glassy 
dynamics in a wider range of colloidal systems dealing with higher order 
multipolar contributions (for instance, quadrupolar moments) or more 
complicated interactions (Janus particles). 

Let us also mention that the present work is also an essential and 
unavoidable first step in a more ambitious program towards a deeper 
study of the non-equilibrium and nonstationary phenomena, such as the 
kinetics of the aging associated with these transitions. As it happens, 
the SCGLE formalism has recently been extended  to describe non-stationary  
non-equilibrium structural relaxation processes, such as aging
or the dependence of the dynamical and structural properties on the 
protocol of fabrication, in liquids constituted by particles interacting 
through non spherical potentials \cite{nescgle7}. The resulting 
non-equilibrium self-consistent generalized Langevin equation theory,  
aimed at describing non-equilibrium phenomena in general, leads in 
particular to a simple and intuitive (but still generic) description of 
the essential behavior of glass-forming dipolar liquids near and beyond 
its dynamical arrest transitions. This renders the 
description of the nonequilibrium processes occurring in a colloidal 
dispersion after an instantaneous temperature quench possible, 
with the most interesting prediction being the aging processes occurring 
when full equilibration is prevented by conditions of dynamical arrest. 
In this regard, the development of the arrest diagram presented in this 
contribution, and its validation by independent results obtained with 
the assistance molecular dynamics simulations constitutes a crucial step 
in developing a full description of dynamical arrest in dipolar liquids. 

Finally, we expect that the results of this work might serve to locate 
and reinterpret previous results dealing with both equilibrium and arrested 
dynamics in dipolar fluids, and as a benchmark for future tests in similar 
models with competitive dynamics arising from distinct degrees of freedom. 
The information provided in this work might be also relevant for the 
rational design of technologically important materials based in ferrofluids, 
considering dipolar systems as prototypical models. Our work is part of 
a long term investigation meant to provide insight in the non-equilibrium 
behavior of various anisotropic systems, such as suspensions of particles 
with higher order multipolar moments, and more complicated interactions, 
for instance, Janus suspensions. Both characterizations will be reported 
in further communications elsewhere.

\begin{acknowledgments}
L.F.E.A acknowledges financial support from the German Academic 
Exchange Service (DAAD) through the DLR-DAAD programme under grant 
No. 212. M.M.N., R.C.P. and G.P.A.  acknowledge the financial 
support provided by the Consejo Nacional de Ciencia y Tecnolog\'{\i}a 
(CONACYT, M\'exico) through grants Nos. 237425 and 287067, 
No. 242364, No. FC-2015-2-1155, and No. LANIMFE-294155. 
The authors thank to Centro de Investigaciones y Estuidos Avanzados 
(CINVESTAV, M\'exico) for the use of the computing clusters
KUKULKAN and ABACUS. 
\end{acknowledgments}

\appendix

\section{Mean square deviations and rotational diffusion coefficients}\label{rotational_msds}

To extract rotational diffusion coefficients from saturating MSDs one can 
start with the solution of Fick's equation for the orientational 
microscopic density, $\rho(\theta,\phi,t)$, on a spherical surface

     \begin{equation}
     \frac{\partial}{\partial t} \rho(\theta, \phi, t)
     = D_{R} \nabla_{s}^2 \rho(\theta, \phi, t),
     \end{equation}

\noindent where $\nabla_s^2$ is the Laplace-Beltrami operator acting over
the surface of the sphere. For an initial condition given by a delta
function on $\theta = 0$ (the north pole), the solution found by
Perrin \cite{perrin-1928} reads 

    \begin{equation}
    \rho(\theta, t) = \sum_{l = 0}^{\infty}
    \frac{2 l + 1}{4 \pi} P_l(\cos \theta)
    e^{-l(l + 1) D_R t},
    \end{equation}

\noindent with $P_l(\cos \theta)$ being Legendre polynomials.
Using this solution, one can calculate the mean square deviation in the
polar angle, $W_{\theta}(t) \equiv \langle \theta^2 (t) \rangle$,
via integration

    \begin{eqnarray}
    W_{\theta}(t)
    &=& \sum_{l = 0}^\infty \frac{2 l + 1}{2}\,
    e^{-l(l + 1) D_R t}
    \int_0^\pi \theta^2
    P_l(\cos \theta) \sin \theta \,  d \theta\\ \nonumber
    &=& a_0 + a_1 e^{-2 D_R t} + a_2 e^{-6 D_R t}
    + a_3 e^{-12 D_R t} + ...
    \end{eqnarray}

\noindent where the coefficients $a_i$ ($i=0,1,2,...$) are easily
calculated, and with the first two being $a_0 = (\pi^2 - 4)/2$ and
$a_1 = -3 \pi^2/8$. Using this result, one may write

    \begin{equation}
     \log(a_0 - W_{\theta}(t))
    =  \log |a_1| -2 D_R t +
    \log \left(1 + b_2 e^{-4 D_R t} + ... \right).
    \end{equation}


\noindent with $b_i$, $i = 2, 3 ...$ being simple geometrical constants
given in terms of $a_i$. Since terms of the form $e^{-n D_Rt}$
decay very fast with increasing $n$, one can make the approximation

    \begin{equation}
    \log(a_0 - W_{\theta}(t))
    \approx  \text{const.} -2 D_R t.
    \label{diff-rot-theta}
    \end{equation}

Using the orthogonality of the Legendre polynomials, Perrin also
found a closed form for $W_{\sin\theta} (t) \equiv \langle \sin^2
\theta(t) \rangle$, which reads

    \begin{equation}
    W_{\sin \theta}(t) = \frac{2}{3} \left(
    1 - e^{-6 D_R t} \right).
    \end{equation}
These two forms can be used, in principle, to get a well-defined value
for the rotational diffusion coefficient. In particular, using
the previous equation, one can write
    \begin{equation}
    \log \left(1 - \frac{3}{2}  W_{\sin \theta}(t) \right) = -6 D_R t.
    \label{diff-rot-sin-theta}
    \end{equation}
One should take into account, however, that for numerical work the
usefulness of eqs. \eqref{diff-rot-theta} and \eqref{diff-rot-sin-theta}
is severely limited by the fact that, as $t$ grows, the argument of the
log function gets very close to zero and the noise overwhelms the signal.
Yet, with good statistics, one may get a fair estimation of the value of
$D_R$. 

\section{Radial distribution function}\label{appendix_gofr}

\begin{figure}
\center
{\includegraphics[width=2.5in, height=2.5in]{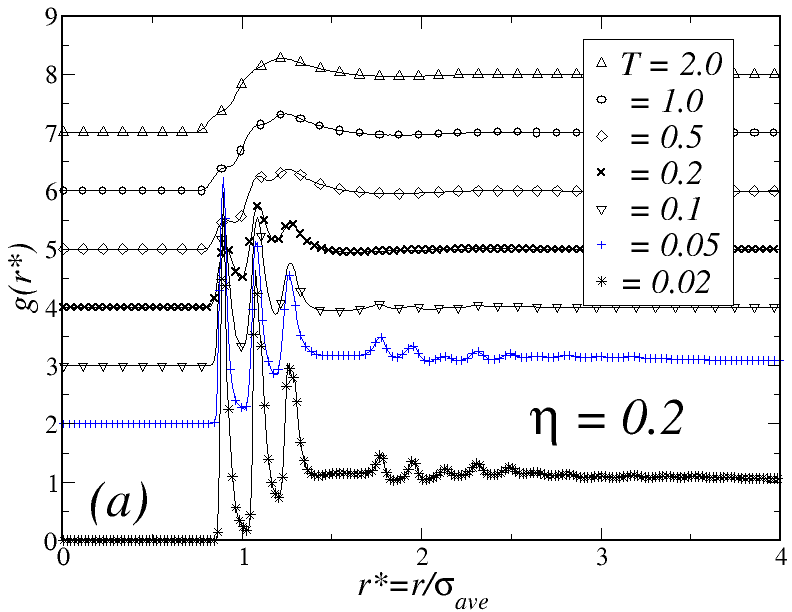}
\includegraphics[width=2.5in, height=2.5in]{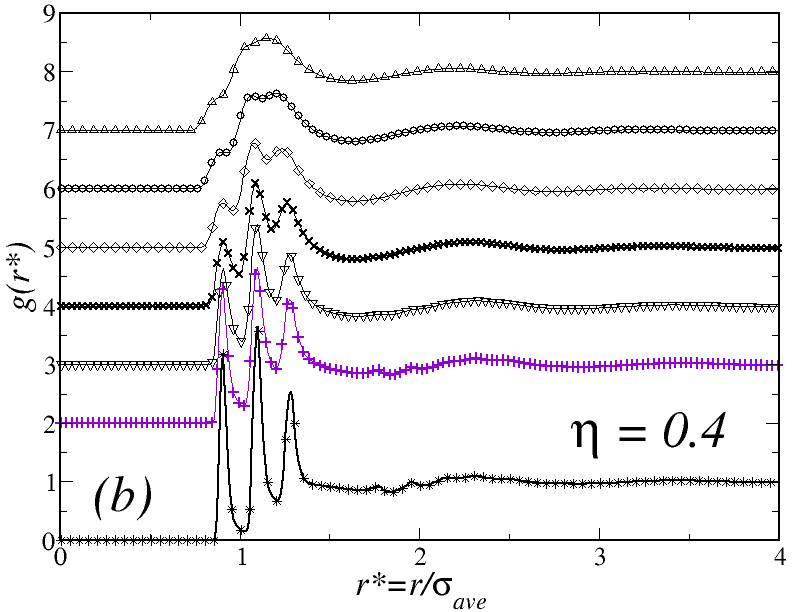}\\
\includegraphics[width=2.5in, height=2.5in]{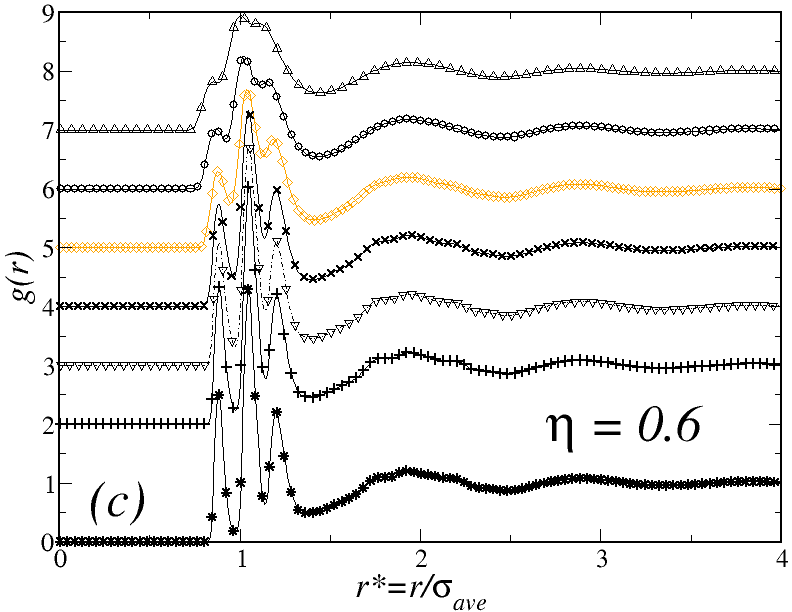}
\includegraphics[width=2.5in, height=2.5in]{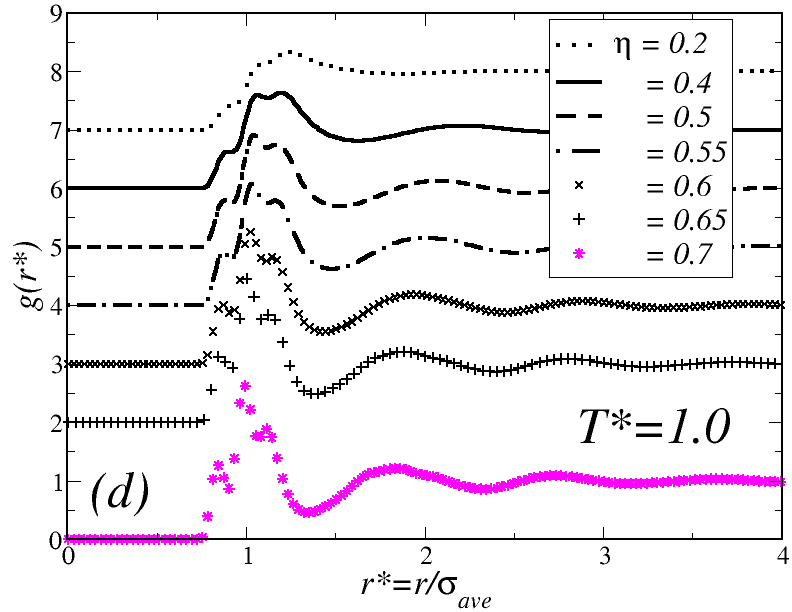}}
\caption{(Color online) Evolution of the radial distribution function, 
$g(r)$ along the three ischores \emph{(a)} $\eta=0.2$ (\emph{Path 1}), 
\emph{(b)} $\eta=0.4$ (\emph{Path 2}), \emph{(c)} $\eta=0.6$ (\emph{Path 3}),
and along the isotherm, $T^*=1$ (\emph{Path 4}).}
\label{gofr}
\end{figure}

Besides the analysis of the dynamics, we have monitored the structural 
behavior of the simulated fluid, represented by the radial distribution 
function $g(r)$, and along the four \emph{Paths} considered. The results 
are shown in Fig. \ref{gofr} (data are shifted in the vertical axis for 
clarity). 

At the isochore $\eta=0.2$ (\emph{Path 1}, Fig. \ref{gofr}(\emph{a})) 
and high temperatures, $g(r)$ exhibits a single broad peak centered at 
$r\approx1.2\sigma_{ave}$, beyond which $g(r)$ attains its asymptotic 
unit value. 
Upon cooling, this first diffuse peak evolves and reveals that is in 
reality the superposition of the nearest-neighbor peaks of $g_{11}(r)$, 
$g_{12}(r)$, and $g_{22}(r)$, separated from each other by approximately 
$\pm\sigma_{ave}/6$.
At $T^*=0.05$ and below, the amplitude of the three peaks become 
noticeable higher. At the same time, one observes the emergence of an 
additional train of smaller peaks, now centered at 
$r\approx2\sigma_{ave}$ representing the second-nearest neighbor shell. 
The various individual peaks in this train correspond to the  
combinations of small and large particles in an approximately linear 
configuration. This may reflect the tendency of the dipolar particles 
to associate in linear trimers and small chains. 

By increasing the volume fraction to $\eta=0.4$, the structure evolves 
to that of a liquid (\emph{Path 2}, Fig. \ref{gofr}\emph{(b)}),
characterized by the emergence of a main peak and followed by other
well-separated, but smaller, secondary peaks. Importantly, the small
oscillations around $r\approx2\sigma_{ave}$ observed in $g(r)$ along 
\emph{Path 1} and at low temperatures disappear. For clarity in the 
comparison, we do not show the behavior of $g(r)$ at the state points 
($\eta=0.4,T^*=0.07$) and ($0.4,T^*=0.03$).

At both the isochore $\eta=0.6$ (\emph{Path 3}, Fig. \ref{gofr}\emph{(c)}) 
and the isotherm $T^*=1$ (\emph{Path 4}, Fig. \ref{gofr}\emph{(d)}), $g(r)$ 
presents the  typical evolution in a dense liquid with short ranged 
repulsion.  

\section{SCGLE theory for Brownian fluids of non-spherical particles}\label{theory}

The SCGLE theory of colloid dynamics and dynamical arrest has been 
previously extended for the description of Brownian fluids constituted 
by particles interacting through non-spherical potentials.
For more details the reader is referred to \cite{Elizondo-PRE-2014}.
In its simplest version, the SCGLE formalism provides a closed
set of time evolution equations for the diagonal components
($l=l'$, $m=m'$) $F_{lm}(k,t)$ and $F^S_{lm}$, of the ISFs defined in 
Eqs. \eqref{isfs} and \eqref{F_self_formula}. 
Written in the Laplace space, such equations read

\begin{equation}
F_{lm}
(k,z)=\frac{S_{lm}(k)}{z+\displaystyle\frac{k^2D_T^0S^{-1}_{lm}(k)}
{1+\Delta\zeta_T^*(z)\lambda^{(lm)}_T(k)}+\frac{l(l+1)D_R^0S^{-1}_{lm}(k)}
{1+\Delta\zeta_R^*(z)\lambda^{(lm)}_R(k)}}\label{flmqframe1}
\end{equation}
and
\begin{equation}
F_{lm}^S (k,z)=\frac{1}{z+\displaystyle\frac{k^2D_T^0}{1+
\Delta\zeta_T^*(z)\lambda^{(lm)}_T(k)}+\frac{l(l+1)D_R^0}
{1+\Delta\zeta_R^*(z)\lambda^{(lm)}_R(k)}}.\label{flmsqframe2}
\end{equation}

\noindent where, $D_R^0$ is the rotational \emph{free}-diffusion
coefficient, and $D^0_T$ is
the center-of-mass translational \emph{free}-diffusion coefficient,
whereas the functions $\lambda^{(lm)}_T(k)$ and $\lambda^{(lm)}_R(k)$
are defined as $\lambda^{(lm)}_T(k)=1/[1+( k/k_{c}) ^{2}]$ and, for
simplicity, $\lambda^{(lm)}_R(k)=1$, where $k_c=\alpha \times k_{max}$,
with $k_{max}$ being the position of the main peak of $S_{00}(k)$ and
$\alpha =1.305$.
This ensures that, for radially-symmetric interactions, one recovers 
the original theory describing liquids of soft and hard spheres.

On the other hand, within a well defined set of approximations 
(discussed in appendix A of Ref. \cite{Elizondo-PRE-2014}) the 
functions $\Delta\zeta_{\alpha}^*(t)$ ($\alpha=T,R$) may be written 
as

\begin{eqnarray}
\Delta\zeta^*_T(t) =\frac{1}{3}\frac{D_T^0}{(2\pi)^3n}\int
d\mathbf{k}
k^2\sum_{l}\left[{2l+1}\right]\left[1-S^{-1}_{l0}(k)\right]^2
F^S_{l0}(k;t)F_{l0}(k;t)\label{zetart}
\end{eqnarray}

\noindent and

\begin{eqnarray}
\Delta\zeta^*_R(t)=\frac{1}{2}\frac{D_R^0}{(2\pi)^3}
\frac{n}{4}\frac{1}{(4\pi)^2}\int
d\mathbf{k}\sum_{l,m}\left[2l+1\right]h^2_{l0}
(k)\left[A_{l;0m}\right]^2\left[S^{-1}_{lm}(k)\right]^2
F^S_{lm}(k;t)F_{lm}(k;t)\label{zetarqq}
\end{eqnarray}
where $h_{lm}(k)$ denotes the diagonal \textit{k}-frame
projections of the total correlation function
$h(\mathbf{k},\bm{\Omega},\bm{\Omega}')$, i.e., $h_{lm}(k)$ is
related to $S_{lm}(k)$ by $S_{lm}(k)=1+(n/4\pi) h_{lm}(k)$, and
$n=N/V$ is the number density. Finally,
$A_{l;mm'}\equiv\left[C_{lm}^+\delta_{m+1,m'}+C_{lm}^-\delta_{m-1,m'}\right]$
and $C_{lm}^{\pm}\equiv\sqrt{(l\mp m)(l\pm m +1)}$.

The closed set of coupled equations (\ref{flmqframe1})-(\ref{zetarqq})  
constitute the equilibrium non spherical version of the SCGLE theory, 
whose solution provides the full wave vector and time dependence of the 
dynamic correlation functions $F_{lm}(k;t)$ and $F^S_{lm}(k;t)$ and of the 
memory functions $\Delta\zeta^{*}_{\alpha}(t)$. These equations may be
numerically solved using standard methods once the projections
$S_{lm}(k)$ of the static structure factor are provided. Under
some circumstances, however, one may only be interested in
identifying and locating the regions in state space of a given system 
that correspond to the various possible ergodic or non ergodic phases,
involving the dynamical arrest of the dynamics of the translational and 
orientational degrees of freedom. For this purpose it is possible to derive 
from the full SCGLE equations the so-called bifurcation equations, i.e.,
the equations for the long-time stationary solutions of equations
(\ref{flmqframe1})-(\ref{zetarqq}). These are written in terms of
the so-called non-ergodicity parameters, defined as

\begin{equation}
f_{lm}(k)\equiv \lim_{t \to
\infty}\frac{F_{lm}(k;t)}{S_{lm}(k)},\label{fc0}
\end{equation}

\begin{equation}
f^S_{lm}(k)\equiv \lim_{t \to
\infty}F^S_{lm}(k;t),\label{fs0}
\end{equation}
and
\begin{equation}
\Delta\zeta_{\alpha}^{*(\infty)} \equiv \lim_{\tau \to \infty}
\Delta\zeta_{\alpha}^{*}(t), \label{dz0}
\end{equation}
with $\alpha= T, R$. The
simplest manner to determine these asymptotic solutions is to take
the long-time limit of Eqs. (\ref{flmqframe1})-(\ref{zetarqq}),
leading to a system of coupled equations for $f_{lm}(k)$, $f^S_{lm}(k)$, and
$\Delta\zeta_{\alpha}^{*(\infty)}$.

It is not difficult to show that the resulting equations  can be
written as
\begin{equation}
f_{lm}(k)=\frac{\left[S_{lm}(k)\right]
\lambda_T^{(lm)}(k)\lambda_R^{(lm)}(k)}{S_{lm}(k)\lambda_T^{(lm)}(k)
\lambda_R^{(lm)}(k)+k^2\gamma_T\lambda_R^{(lm)}(k)+l(l+1)\gamma_R\lambda_T^{(lm)}(k)}\label{fc1}
\end{equation}
and
\begin{equation}
f^S_{lm}(k)=\frac{\lambda_T^{(lm)}(k)\lambda_R^{(lm)}(k)}
{\lambda_T^{(lm)}(k)\lambda_R^{(lm)}(k)+k^2\gamma_T\lambda_R^{(lm)}(k)+l(l+1)
\gamma_R\lambda_T^{(lm)}(k)}\label{fs1},
\end{equation}
where the dynamic order parameters $\gamma_T$ and $\gamma_R$,
defined as
\begin{equation}
\gamma_{\alpha}
\equiv\frac{D^0_{\alpha}}{\Delta\zeta_{\alpha}^{*(\infty)}},
\label{parameters}
\end{equation}
are determined from the solution of
\begin{equation}
\frac{1}{\gamma_T}=\frac{1}{6\pi^2n}\displaystyle{\int_0^\infty}
dk \, k^4\sum_{l}[2l+1]\left[1-S^{-1}_{l0}(k)\right]^2
S_{l0}(k)f^S_{l0}(k)f_{l0}(k),\label{gat}
\end{equation}
and
\begin{equation}
\frac{1}{\gamma_R}=\frac{1}{16\pi^2n}\displaystyle{\int_0^\infty}
dk
k^2\sum_{lm}[2l+1][S_{l0}(k)-1]^2S^{-1}_{lm}(k)f^S_{lm}(k)f_{lm}(k)A_{l;0m}^2
.\label{gar}
\end{equation}

Fully ergodic states are thus described by the condition that the 
non-ergodicity parameters (i.e., $f_{lm}(k)$, $f^S_{lm}(k)$, and
$\Delta\zeta_{\alpha}^{*(\infty)}$) are all zero, so the
dynamic order parameters $\gamma_T$ and $\gamma_R$ are both
infinite. Any other possible solution of these bifurcation
equations indicate total or partial loss of ergodicity. Thus,
$\gamma_T$ and $\gamma_R$ finite indicate full dynamic arrest
whereas $\gamma_T$ finite and $\gamma_R=\infty$ corresponds to the
mixed state in which the translational degrees of freedom are
dynamically arrested but not the orientational degrees of freedom.

\section{Determination of the static structure factor projections, 
$S_{lm}(k)$, for a dipolar soft sphere fuid.}\label{append_msa}

As mentioned above, the solution of Eqs. \eqref{flmqframe1}-\eqref{zetarqq} 
and \eqref{gat}-\eqref{gar} requires the previous determination of the 
spherical harmonics projections $S_{lm}(k)$, of the static structure 
factor of the system under consideration, or equivalently, 
the projections, $c_{lm}(k)$, of the direct correlation function. 
Both quantities are related through the Ornstein-Zernike (OZ) equation
$S_{lm}(k)=\left[1-\displaystyle{\frac{n}{4\pi}c_{lm}(k)}\right]^{-1}$.

For this, one starts by identifying the system of interest, which thus 
require the specific form of the two-particle potential of interaction, 
$U^{ab}(\mathbf{r}_{ab},\hat{\bm{\mu}}_a,\hat{\bm{\mu}}_b)$. 
To represent the simulated system, such pairwise potential should possess 
the generic form  

\begin{equation}
U^{ab}(\mathbf{r}_{ab},\hat{\bm{\mu}}_a,\hat{\bm{\mu}}_b)
=U_{REP}^{ab}(\mathbf{r}_{ab})+
U^{ab}_{DIP}(\mathbf{r}_{ab},\hat{\bm{\mu}}_a,\hat{\bm{\mu}}_b)\label{generic}
\end{equation},
 
\noindent with $U_{REP}^{ab}(\mathbf{r}_{ab})$ describing a short-ranged 
repulsive potential and 
$U^{ab}_{DIP}(\mathbf{r}_{ab},\hat{\bm{\mu}}_a,\hat{\bm{\mu}}_b)$
an anisotropic dipolar contribution. 
This generic representation includes as a particular case the dipolar 
hard-sphere model (DHS), where $U_{REP}^{ab}$ is given by the 
discontinuous hard-sphere potential and $U^{ab}_{DIP}$ by the dipole-dipole 
potential, but it also includes other possible systems. 
For instance, under some circumstances it may be necessary to consider some 
degree of softness in the purely repulsive part, just as it happens in the 
case of the Stockmayer potential \cite{stockmayer}, which replaces 
$U_{REP}^{ab}$ by a Lennard-Jones potential. In general, any such departure 
from the DHS reference potential will destroy the analytical convenience 
provided by the solution of Wertheim for the Mean Spherical Model (MSM) 
\cite{wertheim} which renders the calculation of the functions $c_{lm}(k)$ 
straightforward.

A simplified version of the Stockmayer potential which allows 
to exploit the analytical simplicity of Wertheim's solution, consist 
in replacing $U_{REP}^{ab}$ by the WCA potential of Eq. 
\eqref{wcapotential}. In this manner, the generic potential of Eq. 
\eqref{generic} can be written as the following soft-sphere 
dipolar potential

\begin{equation}
U^{ab}_{SSD}(\mathbf{r}_{ab},\hat{\bm{\mu}}_a,\hat{\bm{\mu}}_b)
=U_{WCA}^{ab}(\mathbf{r}_{ab})+
\frac{\epsilon^{ab}_{\text{\tiny{DIP}}}}{(r_{ab}/\sigma)^3}
[\bm{\hat{\mu}}_a\cdot\bm{\hat{\mu}}_b
-3(\bm{\hat{\mu}}_a\cdot\hat{\mathbf{r}}_{ab})
(\bm{\hat{\mu}}_b\cdot\hat{\mathbf{r}}_{ab})]\label{ssd}
\end{equation},

\noindent where, in the case of a monodisperse system, 
$\epsilon^{ab}_{\small{DIP}}=\mu^2/\sigma^3$ is the energy 
scale of the dipolar contribution. A further simplification arises 
when one considers $\epsilon^{ab}_{WCA}=\epsilon^{ab}_{DIP}$

In the spirit of the WCA treatment of soft-core interactions \cite{hansen},
we may assume that the properties of the soft-core dipolar 
potential of Eq. \ref{ssd} can be approximated by those of an effective DHS 
potential $U^{eff}_{DHS}(\mathbf{r}_{ab},\hat{\bm{\mu}}_a,\hat{\bm{\mu}}_b)$
with a state-dependent effective diameter $\sigma^{eff}(n,T)$, \emph{i.e.},
by a system with a potential of interaction 

\begin{equation}
U^{eff}_{DHS}(\mathbf{r}_{ab},\hat{\bm{\mu}}_a,\hat{\bm{\mu}}_b)
=U^{eff}_{HS}(r;\sigma^{eff})+\frac{\epsilon^{ab(eff)}_{DIP}}{(r/\sigma^{eff})^3}
[\bm{\hat{\mu}}_a\cdot\bm{\hat{\mu}}_b
-3(\bm{\hat{\mu}}_a\cdot\hat{\mathbf{r}}_{ab})
(\bm{\hat{\mu}}_b\cdot\hat{\mathbf{r}}_{ab})]
\end{equation}

\noindent where $\epsilon^{ab(eff)}_{DIP}$ is defined as 
$\epsilon^{ab(eff)}_{DIP}\equiv \lambda{-3}\epsilon^{ab}_{DIP}$, and with 
$\lambda\equiv\sigma^{eff}(n,T)/\sigma$. The similarity between the WCA 
and the HS potentials leads to the additional simplification that
$\sigma_{HS}(n,T)$ becomes $n$-independent, and given by the "blip function" 
approximation \cite{hansen}.

Therefore, within these assumptions, the direct correlation function 
$c_{SSD}((\mathbf{r}_{ab},\hat{\bm{\mu}}_a,\hat{\bm{\mu}}_b);n,T)$ of a soft 
dipolar system with a potential defined by Eq. \ref{ssd} can be approximated 
by the direct correlation function of an effective DHS system,  
$c^{eff}_{DHS}(\mathbf{r}_{ab},\hat{\bm{\mu}}_a,\hat{\bm{\mu}}_b;n,T)$.
In dimensionless units, this approximation reads

\begin{equation}
c_{SSD}(\mathbf{r}/\sigma,\hat{\bm{\mu}}_1,\hat{\bm{\mu}}_2;n^*,T^*)
\approx
c_{DHS}(\mathbf{r}/\lambda\sigma,\hat{\bm{\mu}}_1,\hat{\bm{\mu}}_2;
\lambda^3n^*,\lambda^{-3}T^*).
\end{equation}

Clearly, introducing again the MSA  for the calculation of the right side 
of this equation restores the analytical simplicity of Wertheim's solution
by means of a simple re-scaling procedure. From here, the determination of 
the projections $c_{lm}(k)$ is straightforward (see, for instance, appendix 
E of Ref. \cite{schilling-PRE-1997})

\end{document}